\documentclass{sigchi}

\let\citep\cite
\let\citet\cite

\def\tightlist{} 

\CopyrightYear{2020}
\setcopyright{rightsretained}

\doi{https://doi.org/10.1145/3313831.3376672}
\isbn{978-1-4503-6708-0/20/04}
\conferenceinfo{CHI'20,}{April 25--30, 2020, Honolulu, HI, USA}
\acmPrice{\$15.00}

\toappear{\scriptsize Permission to make digital or hard copies of part or all of this work for personal or classroom use is granted without fee provided that copies are not made or distributed for profit or commercial advantage and that copies bear this notice and the full citation on the first page. Copyrights for third-party components of this work must be honored. For all other uses, contact the owner/author(s). \\ 
{\emph{CHI '20, April 25--30, 2020, Honolulu, HI, USA.} } \\
Copyright is held by the author/owner(s). \\ 
ACM ISBN 978-1-4503-6708-0/20/04. \\
http://dx.doi.org/10.1145/3313831.3376672}

\usepackage{balance}       
\usepackage{graphics}      
\usepackage[T1]{fontenc}   
\usepackage{txfonts}
\usepackage{mathptmx}
\usepackage[pdflang={en-US},pdftex]{hyperref}
\usepackage{color}
\usepackage{booktabs}
\usepackage{textcomp}

\usepackage{microtype}        
\usepackage{ccicons}          

\usepackage{todonotes}

\def\plaintitle{`I Just Want to Hack Myself to Not Get Distracted':\\\mbox{ Evaluating Design Interventions for Self-Control on Facebook}}

\def\emptyauthor{}
\def\plainkeywords{Facebook; problematic use; self-control; distraction; ICT non-use; addiction; focus; interruptions}

\makeatletter
\def\url@leostyle{%
  \@ifundefined{selectfont}{
    \def\UrlFont{\sf}
  }{
    \def\UrlFont{\small\bf\ttfamily}
  }}
\makeatother
\def\pprw{8.5in}
\def\pprh{11in}

\definecolor{linkColor}{RGB}{6,125,233}
\hypersetup{%
  pdftitle={\plaintitle},
  pdfauthor={\emptyauthor},
  pdfkeywords={\plainkeywords},
  pdfdisplaydoctitle=true, 
  bookmarksnumbered,
  pdfstartview={FitH},
  colorlinks,
  citecolor=black,
  filecolor=black,
  linkcolor=black,
  urlcolor=linkColor,
  breaklinks=true,
  hypertexnames=false
}
\usepackage{url}

\usepackage{breakurl}

\begin{document}

\title{\plaintitle}

\numberofauthors{1}
\author{%
  \alignauthor{Ulrik Lyngs\textsuperscript{1}, Kai Lukoff\textsuperscript{2}, Petr Slovak\textsuperscript{3}, William Seymour\textsuperscript{1}, Helena Webb\textsuperscript{1}, \linebreak Marina Jirotka\textsuperscript{1}, Jun Zhao\textsuperscript{1}, Max Van Kleek\textsuperscript{1}, Nigel Shadbolt\textsuperscript{1}\\
    \affaddr{\textsuperscript{1}Department of Computer Science, University of Oxford, UK}\\
    \email{\{ulrik.lyngs, william.seymour, helena.webb, marina.jirotka,\\jun.zhao, max.van.kleek, nigel.shadbolt\}@cs.ox.ac.uk}\\
    \vspace{0.5mm}
    \affaddr{\textsuperscript{2}Human Centered Design \& Engineering, University of Washington, Seattle, US,}
    \email{kai1@uw.edu}\\
    \vspace{0.5mm}
    \affaddr{\textsuperscript{3}Department of Informatics, King's College London, UK,}
    \email{petr.slovak@kcl.ac.uk}}
}

\maketitle

\begin{abstract}
Beyond being the world's largest social network, Facebook is for many also one of its greatest sources of digital distraction.
For students, problematic use has been associated with negative effects on academic achievement and general wellbeing.
To understand what strategies could help users regain control, we investigated how simple interventions to the Facebook UI affect behaviour and perceived control.
We assigned 58 university students to one of three interventions: goal reminders, removed newsfeed, or white background (control).
We logged use for 6 weeks, applied interventions in the middle weeks, and administered fortnightly surveys.
Both goal reminders and removed newsfeed helped participants stay on task and avoid distraction.
However, goal reminders were often annoying, and removing the newsfeed made some fear missing out on information.
Our findings point to future interventions such as controls for adjusting types and amount of available information, and flexible blocking which matches individual definitions of `distraction'.
\end{abstract}




\enlargethispage{3\baselineskip}
\hypertarget{introduction}{%
\section{Introduction}\label{introduction}}

Research on `Problematic Facebook Use' (PFU) has investigated correlations between Facebook use and negative effects on outcomes such as level of academic achievement \citep{guptaInclassDistractionsRole2016} and subjective wellbeing \citep{marinoComprehensiveMetaanalysisProblematic2018, marinoAssociationsProblematicFacebook2018}.
Here, a cross-cutting finding is that negative outcomes are associated with subjective difficulty at exerting self-control over use, as well as specific use patterns including viewing friends' wide-audience broadcasts rather than receiving targeted communication from strong ties \citep{burkeRelationshipFacebookUse2016, marinoComprehensiveMetaanalysisProblematic2018}.

Much of this work has focused on self-control over Facebook use in student populations \citep{al-dubaiAdverseHealthEffects2013, khumsriPrevalenceFacebookAddiction2015, kocFacebookAddictionTurkish2013}, with media multitasking research finding that students often give in to use which provides short-term `guilty pleasures' over important, but aversive academic tasks \citep{rosenFacebookTextingMade2013, xuMediaMultitaskingWellbeing2016, meierFacebocrastinationPredictorsUsing2016}.
In the present paper, we present a mixed-methods study exploring how two interventions to Facebook --- goal reminders and removing the newsfeed --- affect university students' patterns of use and perceived control over Facebook use.
To triangulate self-report with objective measurement, our study combined usage logging with fortnightly surveys and post-study interviews.

\enlargethispage{3\baselineskip}

We found that both interventions helped participants stay on task and use Facebook more in line with their intentions.
In terms of usage patterns, goal reminders led to less scrolling, fewer and shorter visits, and less time on site, whereas removing the newsfeed led to less scrolling, shorter visits, and less content 'liked'.
However, goal reminders were often experienced as annoying, and removing the newsfeed made some participants fear missing out on information.
After the study, participants suggested a range of design solutions to mitigate self-control struggles on Facebook, including controls for filtering or removing the newsfeed, reminders of time spent and use goals, and removing features that drive engagement.
As an exploratory study, this work should be followed by confirmatory studies to assess whether our findings replicate, and how they may generalise beyond a student population.

\hypertarget{related-work}{%
\section{Related work}\label{related-work}}

\hypertarget{struggles-with-facebook-use}{%
\subsection{Struggles with Facebook use}\label{struggles-with-facebook-use}}

Whereas many uses of Facebook offer important benefits, such as social support, rapid spread of information, or facilitation of real-world interactions \citep{ryanUsesAbusesFacebook2014}, a substantial amount of research has focused on negative aspects \citep{marinoComprehensiveMetaanalysisProblematic2018}.
For example, studies have reported correlations between patterns of Facebook use and lower academic achievement \citep{rouisImpactFacebookUsage2011, wangContextCollegeStudents2018}, low self-esteem, depression and anxiety \citep{labragueFacebookUseAdolescents2014}, feelings of isolation and loneliness \citep{al-dubaiAdverseHealthEffects2013}, and general psychological distress \citep{chenSharingLikingCommenting2013}.
Such `Problematic Facebook Use' (PFU) has been studied under various names (including `Facebook dependence' \citep{wolniczakAssociationFacebookDependence2013} and `Facebook addiction'\citep{andreassenDevelopmentFacebookAddiction2012}), but a recent review summarised a common definition across papers as `problematic behaviour characterised by addictive-like symptoms and/or self-regulation difficulties related to Facebook use leading to negative consequences in personal and social life' \citep{marinoComprehensiveMetaanalysisProblematic2018}.

A large number of studies have in turn correlated measures of PFU with patterns of use and personality traits. Here, researchers often distinguish between use that is more `active' (creating content and communicating with friends) and use that is more `passive' (consuming content created by others without actively engaging), with the former being linked to more positive correlates of subjective wellbeing \citep{burkeSocialNetworkActivity2010, ellisonBenefitsFacebookFriends2007, gersonPassiveActiveFacebook2017, grieveFacetofaceFacebookCan2013} and the latter to more negative \citep{krasnovaEnvyFacebookHidden2013, verduynPassiveFacebookUsage2015}.

Moreover, most studies have found that `problematic users' tend to spend more time on Facebook \citep{marinoComprehensiveMetaanalysisProblematic2018}, including a recent study by researchers at Facebook with direct access to server logs: users who experienced their use as problematic (i.e., reported negative impact on sleep, relationships, or work/school performance, plus lack of control over use) spent more time on the platform, especially at night, as well as more time looking at profiles and less time browsing the newsfeed, and were more likely to deactivate their accounts \citep{chengUnderstandingPerceptionsProblematic2019}.

Depending on the specific tools and thresholds used for assessing use as `problematic', prevalence estimates vary widely, from 3.1\% in a representative sample of US users \citep{chengUnderstandingPerceptionsProblematic2019} to 47\% in a study of Malaysian university students (\citep{jafarkarimiFacebookAddictionMalaysian2016} see also \citep{banyaiProblematicSocialMedia2017, khumsriPrevalenceFacebookAddiction2015, wolniczakAssociationFacebookDependence2013}).
The upper bounds of such estimates suggest that, at least at a mild levels, it is very common for people to struggle with using Facebook in accordance with their goals \citep{guedesInternetAddictionExcessive2016}.
This is supported by studies of multitasking and media use finding that people very often perceive their use of digital media to be in conflict with other important goals (61.2\% of use occurrences in an experience sampling study by Reinecke and Hofmann \citep{reineckeSlackingWindingExperience2016}), and that Facebook in particular is one of the most common sources of media-induced procrastination \citep{rosenFacebookTextingMade2013, xuMediaMultitaskingWellbeing2016}.

\enlargethispage{3\baselineskip}
\hypertarget{interventions-and-digital-self-control-tools}{%
\subsection{Interventions and digital self-control tools}\label{interventions-and-digital-self-control-tools}}

Catering to users struggling with self-control over digital device use, a growing niche exists for `digital self-control' tools on online stores for apps and browser extensions.
Such tools promise to support user self-control through interventions such as removing distracting element from websites, tracking and visualising use, or rewarding intended behaviour \citep{lyngsSelfControlCyberspaceApplying2019}.
In particular, many browser extensions focus on adjusting \emph{Facebook} in ways intended to help self-control, for example by removing the newsfeed \citep{jdevNewsFeedEradicator2019} or hiding numerical metrics such as like count \citep{grosserFacebookDemetricator2019}.

No studies have assessed how interventions found in these tools may alleviate self-control struggles on Facebook.
However, some recent studies have investigated how temporarily deactivating or not logging in to Facebook affect subjective wellbeing \citep{allcottWelfareEffectsSocial2019, mosqueraEconomicEffectsFacebook2019, tromholtFacebookExperimentQuitting2016, vanmanBurdenOnlineFriends2018}.
The findings from these studies have largely been in agreement, with Allcott et al. \citep{allcottWelfareEffectsSocial2019} the largest to date:
in a study where 580 participants were randomly assigned to deactivate their accounts for four weeks, and compared to 1,081 controls, Facebook deactivation increased offline activities (including socialising with family and friends and watching TV) and subjective wellbeing, and decreased online activity (including other social media than Facebook).
Moreover, Facebook deactivation caused a large and persistent reduction in Facebook use after the experiment.

For many users, however, deactivating or deleting their Facebook account presents too tall a barrier to action for tackling problematic use.
Most users have more targeted non-use goals than ``abstinence'', such as reducing time scrolling the newsfeed (but not time posting in a university social group), or reducing time spent on Facebook during final exams (but not during vacations, cf. \citep{chengUnderstandingPerceptionsProblematic2019, wangContextCollegeStudents2018}).
Some existing research similarly supports positive effects on wellbeing of targeted non-use, including research on active versus passive social media use \citep{burkeRelationshipFacebookUse2016, hinikerMyTimeDesigningEvaluating2016, verduynPassiveFacebookUsage2015}.
Therefore, investigating interventions found in digital self-control tools for Facebook presents an exciting research opportunity, as they represent less extreme measures than deactivation that may have positive effects.

\enlargethispage{3\baselineskip}
\hypertarget{overview-of-study}{%
\section{Overview of study}\label{overview-of-study}}

On this background, we set out to study how two interventions found in popular browser extensions for scaffolding self-control on Facebook --- specifically, \emph{adding goal prompts and reminders} and \emph{removing the newsfeed} --- affect patterns of use and perceived control on Facebook among university students.
We designed a mixed-methods study that attempted to address common limitations in related studies:

\begin{itemize}
\item
  Most studies rely on self-reported Facebook use, which complicates interpretation because self-report often correlates poorly with actual use of digital devices \citep{ellisDigitalTracesBehaviour2018, ellisSmartphoneUsageScales2019, orbenScreensTeensPsychological2019, scharkowAccuracySelfReportedInternet2016}.
  Therefore, we combined surveys and interviews with logging of use, to triangulate subjective self-report and objective measurement.
\item
  Nearly all studies, apart from deactivation studies, have used cross-sectional designs, making it very difficult to interpret causality \citep{marinoComprehensiveMetaanalysisProblematic2018}.
  Therefore, we randomly assigned participants to intervention groups and compared an initial baseline to a subsequent intervention as well as post-intervention block.
\end{itemize}

Our choice of interventions is described in the `Pre-study' section below.
Based on existing research on self-control struggles in relation to Facebook use, our research questions were as follows:

\begin{itemize}
\tightlist
\item
  RQ1 (Amount of use): How do goal reminders (C\textsubscript{goal}) or removing the newsfeed (C\textsubscript{no-feed}) impact time spent and visits made?
\item
  RQ2 (Patterns of use): How do goal reminders (C\textsubscript{goal}) or removing the newsfeed (C\textsubscript{no-feed}) impact patterns of use (e.g., passive / active)?
\item
  RQ3 (Control): How do goal reminders (C\textsubscript{goal}) or removing the newsfeed (C\textsubscript{no-feed}) impact perceived control?
\item
  RQ4 (Post-intervention effects): Do the effects (RQ1-3) of goal reminders (C\textsubscript{goal}) or removing the newsfeed (C\textsubscript{no-feed}) persist after interventions are removed?
\item
  RQ5 (Self-reflection): Do the interventions enable participants to reflect on their struggles with Facebook use in ways that might inform the design of more effective interventions?
\end{itemize}

Whereas RQ1-4 follow from the background literature reviewed, RQ5 was a generative research question pointing towards new design solutions.
We did not envision participants being `vessels of truth' in relation to which design interventions would solve their struggles, but were interested in what suggestions the interventions might inspire as design probes.

\section{Methods}

Study materials, anonymised data, and analysis scripts are available via the \href{https://osf.io}{Open Science Framework} at \href{https://osf.io/qtg7h/}{osf.io/qtg7h}.


\hypertarget{pre-study}{%
\subsection{Pre-study}\label{pre-study}}

\subsubsection{Reviewing Facebook self-control tools}
In February 2018, we searched for browser extensions for supporting self-control on Facebook on the Chrome Web store and identified 50 such extensions implementing a range of interventions (see study materials for the list).
Most (36 out of 50) let the user remove or alter distracting elements, with more than half (27/50) specifically hiding the newsfeed (e.g., `Newsfeed Eradicator' \citep{jdevNewsFeedEradicator2019} removes the newsfeed and optionally replaces it with a motivational quote).
Others implemented interventions such as time limits (e.g., setting a daily limit and prompting the user to stop using Facebook or, like \emph{Auto Logout} \citep{avtechlabsAutoLogout2019}, force closing it when the time has passed), goal reminders (e.g., asking the user what she needs to do on Facebook and subsequently providing reminders, \emph{Focusbook} \citep{forstyonahFocusbook2016}) or providing rewards or punishments (e.g., transferring money out of one's bank account if use is above a set limit, \emph{Timewaste Timer} \citep{prettymind.coTimewasteTimer2018}).

\enlargethispage{3\baselineskip}
\subsubsection{Selecting interventions to investigate guided by a dual systems model of self-control}
To categorise these interventions, we relied on a recent review of functionality in digital self-control tools, which grouped their main design features into the types \emph{block/removal}, \emph{self-tracking}, \emph{goal advancement}, and \emph{reward/punishment}, which in turn were mapped to psychological mechanisms in a dual systems model of self-regulation \citep{lyngsSelfControlCyberspaceApplying2019}.
This model distinguishes between behaviour under non-conscious `System 1' control, i.e., when the external environment or internal states trigger habits or instinctive responses; and behaviour that is under conscious `System 2' control, i.e., when goals, intentions, and rules held in working memory trigger behaviour.
For example, a student might check Facebook as the first thing when opening his laptop, because this context triggers a habitual check-in via System 1 control.
Alternatively, the student might open Facebook because he has a conscious goal of messaging a friend.

According to this model---which we return to in the Discussion---`self-control' is the capacity of conscious System 2 control to override System 1 responses when the two are in conflict.
For example, one might have a conscious goal to not check one's phone at the dinner table and a need to use self-control to suppress one's checking habit, in order to align behaviour with this goal (see \citep{lyngsSelfControlCyberspaceApplying2019} for details, cf. \citep{kotabeIntegratingComponentsSelfControl2015}).

\emph{C\textsubscript{no-feed}} {} For our first experimental condition, C\textsubscript{no-feed}, we chose removing the newsfeed, because this was by far the most common approach among the extensions reviewed.
Viewed through the dual systems model lens, removing the newsfeed represents a \emph{block/removal} strategy which scaffolds self-control on Facebook by preventing unwanted System 1 control from being triggered by the newsfeed, and supporting System 2 control by preventing distracting information from crowding out working memory and make the user forget her goal \citep{lyngsSelfControlCyberspaceApplying2019}.

\emph{C\textsubscript{goal}} {} To compare this to a different strategy, we selected a \emph{goal advancement} intervention as a second experimental condition (C\textsubscript{goal}), specifically the one implemented by \emph{Focusbook} \citep{forstyonahFocusbook2016}, which prompts the user to type in their goal when visiting Facebook and then periodically reminds them of this goal.
According to the dual systems model, this scaffolds self-control in a way that is distinct from removing the newsfeed, namely, by keeping the goals the user wishes to achieve present in working memory, thereby enabling System 2 control.
We chose \emph{Focusbook}'s implementation, because it had the largest number of users among the extensions reviewed that implemented alternatives to \emph{block/removal} strategies.

\emph{C\textsubscript{control}} {} In order to control for `demand characteristics' and placebo effects \citep{bootPervasiveProblemPlacebos2013, nicholsGoodSubjectEffectInvestigating2008}, we also included a control condition (C\textsubscript{control}).
In this condition, we changed the background colour of Facebook from light grey to white, which we did not hypothesise to have any significant effect on behaviour or perceived control.

\enlargethispage{3\baselineskip}
\hypertarget{materials}{%
\subsection{Materials}\label{materials}}

\hypertarget{study-conditions}{%
\subsubsection{Study conditions}\label{study-conditions}}

\begin{figure}
\includegraphics[width=1\linewidth]{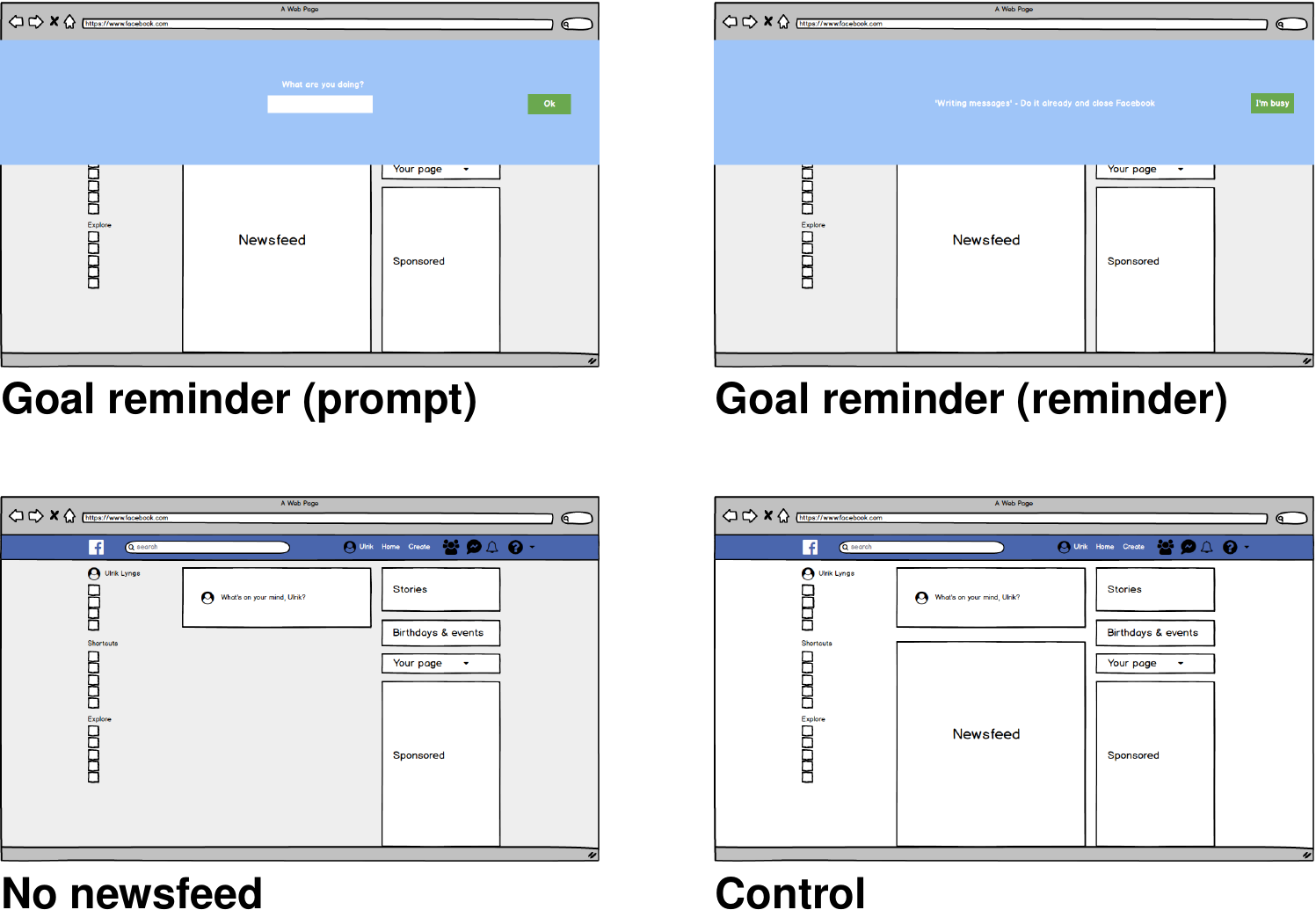} \caption{Mockup of study conditions: C\textsubscript{goal} (adding a goal prompt when visiting the site that every few minutes pops up a reminder), C\textsubscript{no-feed} (removing the newsfeed), and C\textsubscript{control} (white background). See study materials for screenshots.}\label{fig:study-conditions}
\end{figure}

The study conditions are shown in Figure \ref{fig:study-conditions}.\\
We implemented the interventions as Chrome extensions written in JavaScript and CSS:
during the intervention block, the extension script for C\textsubscript{control} turned the background colour of Facebook white.
For C\textsubscript{no-feed}, the extension script hid the webpage elements containing the newsfeed.
For C\textsubscript{goal}, the extension script was a modified version of \emph{Focusbook} (the source code for which is available on GitHub \citep{forstyonahFocusbook2016}), where we forced safe-for-work-mode (i.e., avoiding foul language in reminders) and altered prompts that expressed disapproval to neutral reminders (e.g., changing ``Fine, just tell me why you needed to open Facebook'' to ``Tell me why you needed to open Facebook'').
The extension prompted the user to type in why they opened Facebook when they went to the site, and after 1-3 minutes popped up a reminder of what they typed, along with a snooze button.
Until the snooze button was pressed, the banner containing the prompt slowly expanded to take up more and more screenspace.

\hypertarget{logging-of-use}{%
\subsubsection{Logging of use}\label{logging-of-use}}

Following recent work \citep{wangContextCollegeStudents2018}, we used the open-source browser extension `Research tool for Online Social Environments' (ROSE) \citep{pollerWelcomeROSEResearch2017, pollerInvestigatingOSNUsers2014} to log Facebook use in the Google Chrome browser.
We used this extension to record usage metrics (e.g., timestamps when a browser tab with Facebook was brought in and out of focus, number of clicks) and specific interactions (e.g., viewing a profile, liking content).
To preserve privacy, the extension gave interactions (e.g., content liked) an anonymous identifier in stored data without storing any identifying information about the actual content engaged with.
The ROSE extension was installed on participants' laptop in addition to the extension for their intervention condition.

\hypertarget{surveysinterviews}{%
\subsubsection{Surveys/interviews}\label{surveysinterviews}}

\textbf{Opening survey}:
The opening survey included
(i) demographic information,
(ii) basic information about participants' use of Facebook (when they got an account, devices they use to access the site, prior use of self-control tools for Facebook), and
(iii) two individual difference measures (susceptibility to types of distraction \citep{markEffectsIndividualDifferences2018} and a Big Five personality measure \citep{goslingVeryBriefMeasure2003}).

\textbf{Repeated surveys}:
The survey administered after each study block included three measures:\\
(i) The Passive and Active Facebook Use Measure (PAUM; \citep{gersonPassiveActiveFacebook2017}), which assesses frequency of activities on Facebook.
The measure is factored into the usage dimensions `active social' (items including ``Posting status updates'', ``Chatting on FB chat''), `active non-social' (e.g., ``Creating or RSVPing to events'', ``Tagging photos''), and `passive' (e.g., ``Checking to see what someone is up to'', ``Browsing the newsfeed passively (without liking or commenting on anything)'').\\
(ii) The Multidimensional Facebook Intensity Scale \citep{oroszFourFacetsFacebook2016}, which assesses agreement with statements about Facebook use (e.g., ``I feel bad if I don't check my Facebook daily'') and is factored into the dimensions `persistence', `boredom', `overuse' and `self-expression'.\\
(iii) The Single-Item Self-Esteem Scale \citep{robinsMeasuringGlobalSelfesteem2001}, a commonly used measure of self-esteem in psychological research.

In addition, the survey after the intervention block included items on whether the changes affected perceived control, or how participants accessed Facebook on laptop vs smartphone.

\textbf{Interviews}:
After the study, we conducted semi-structured interviews with all participants.
Main topics probed were
(i) whether the interventions worked as expected,
(ii) how participants experienced the interventions (example question: ``When {[}changes in the participant's condition{]}, what was that like?''),
(iii) what changes participants might wish to make to Facebook to support their intended use (example question: ``If you could build any extension you wanted to change the way Facebook appears and works to make it work better for you, what might you want to do?'').

\textbf{5-month follow-up survey}:
Five months after the study, we sent participants an optional brief survey, assessing whether (and if so, how) the study had led to enduring changes in how they use Facebook.

\enlargethispage{2\baselineskip}
\hypertarget{recruitment}{%
\subsection{Recruitment}\label{recruitment}}

Participants were recruited from colleges at the University of Oxford, using a combination of mailouts, posters, and Facebook posts.
Recruitment materials described the study as a study on `Facebook distraction', investigating `which parts of Facebook distract users, and what might be done about it'.
Recruitment targeted non-first year students aged 18-30, who felt they were `often distracted by Facebook'.
Participation was compensated with a \pounds20 Amazon gift card.

\hypertarget{procedure}{%
\subsection{Procedure}\label{procedure}}

\begin{figure}
\includegraphics[width=1\linewidth]{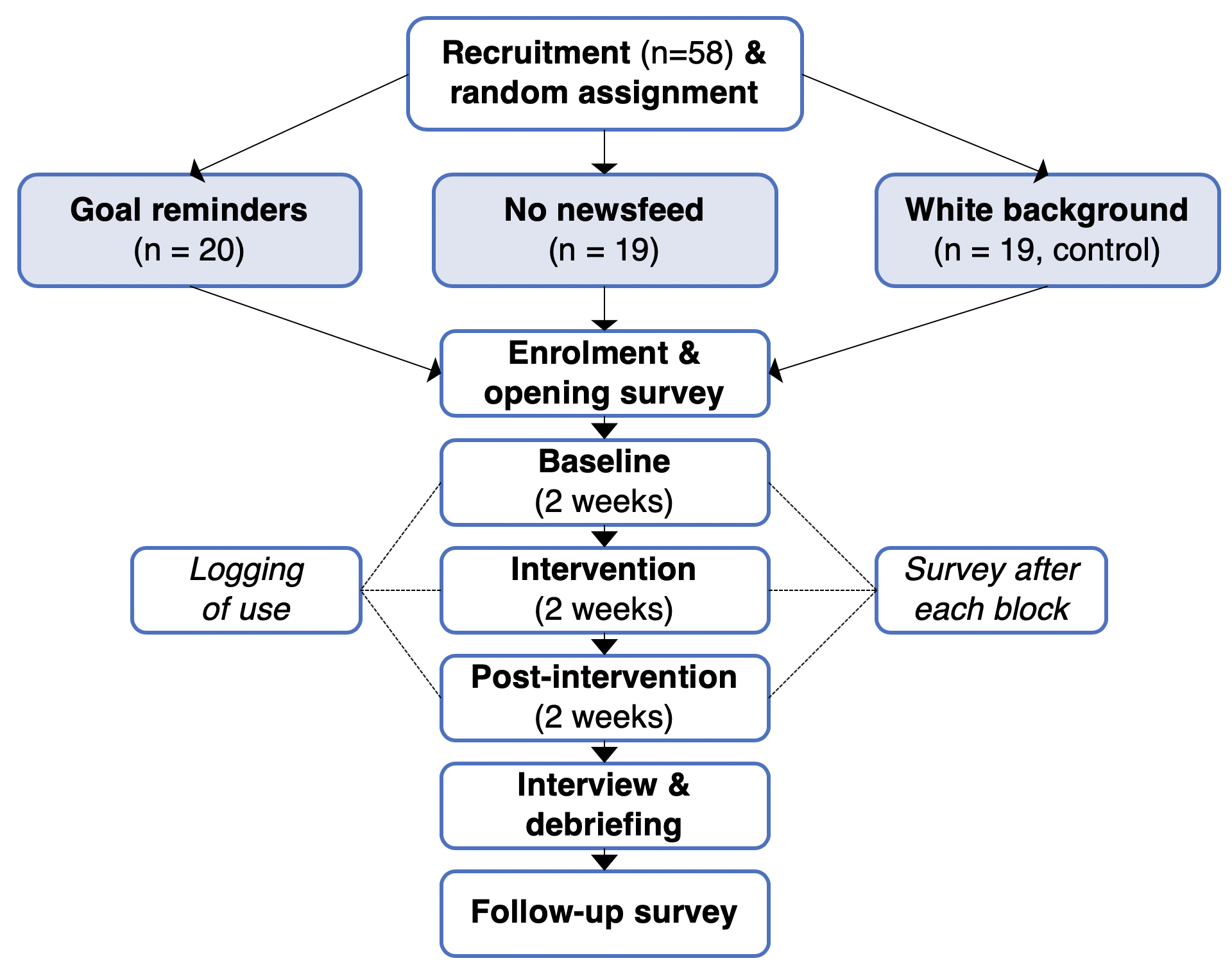} \caption{Flowchart of the study procedure}\label{fig:flowchart}
\end{figure}

\enlargethispage{2\baselineskip}
A flowchart of the study procedure is shown in Figure \ref{fig:flowchart}.\\
Participants were randomly assigned to conditions.
At an initial meeting, participants filled in the opening survey and installed two extensions on their laptop for the Chrome browser:
the ROSE extension for logging use and our extension for modifying Facebook according to their assigned condition.
Participants were instructed to use Chrome whenever they accessed Facebook on their laptop throughout the study period, and informed that the extensions would `anonymously measure how you spend time on the site' and `may change how Facebook appears at some point during the study period'.
The logging period lasted six weeks, grouped into three two-week blocks.
By the end of each block, participants were sent a survey link on Friday at 3pm and a reminder two days later.
The first block served as a baseline, with no changes made to Facebook.
In the second block, interventions were applied from Monday 9am (announced with a pop-up the first time participants visited Facebook) to Monday 9am two weeks later.
The third block served as a new baseline measurement (post-intervention) with Facebook returned to normal.
By the end of this block, a pop-up thanked participants for taking part and directed them to sign up for an interview and debriefing.

A subset of participants (n = 11) began the tracking period one week later than the others.

\hypertarget{data-pre-analysis}{%
\subsection{Data pre-analysis}\label{data-pre-analysis}}

\textbf{Quantitative data}:
On rare occasions, the ROSE extension did not correctly record entries to or exits from Facebook, which resulted in some instances where the calculated duration of active focus on a tab with Facebook was unrealistically long (more than 24 hours in one case).
To handle such instances, we excluded visits longer than one hour when analysing visit durations (144 tab visits out of a total of 120,002).\footnote{See study materials for the precise data processing workflow from raw data to reported results.}

\textbf{Interview transcription and thematic analysis}:
Two of the authors transcribed and conducted thematic analysis of all the interviews and free-text survey responses.
The recordings were iteratively transcribed and analysed using an open-coding approach.
The authors reviewed transcripts and identified emerging codes individually, and regularly discussed emerging codes.

Thematic analysis was conducted in the \href{https://www.dedoose.com}{Dedoose} software; quantitative analyses were conducted in \href{https://www.r-project.org}{R}.

\enlargethispage{4\baselineskip}
\hypertarget{results}{%
\section{Results}\label{results}}

58 students (21 male) took part.
For 8 participants, the intervention failed (on some Windows laptops, security settings prompted participants to turn the extensions off), and 1 participant deactivated his Facebook account during the study.
Survey and logging data from these participants, as well as their interview statements about the interventions, were excluded from analysis.
In addition, 2 participants deleted the ROSE extension before the debriefing - and with it their logged use - and for 1 participant the interview recording device failed.
This left us with survey data from 49 participants, logging data from 47 participants, and interview data from 57 participants for analysis.
Median interview length was 23m 51s (sd = 5m 5s).

In the following, we first report general characteristics of participants and their Facebook use, as well as introductory notes on how interventions were used and perceived. Afterwards, we report results grouped by research question.

\hypertarget{participant-characteristics}{%
\subsection{Participant characteristics}\label{participant-characteristics}}

Participants' median age was 22.5 (min = 19, max = 38) years.
90\% had had a Facebook account for six years or longer, and the median number of Facebook friends was 900 (min = 200, max = 2200).
All participants used Facebook on their laptop, and
96\% also used it on their smartphone. 
On smartphone, most (78\%) used the Facebook and Messenger apps, 8\% used the web browser (instead of the Facebook app) plus the Messenger app, 6\% used only the Messenger app, and 2\% (1 participant) used only the smartphone's web browser to access Facebook.

Most participants (71\%) had never used digital self-control tools for Facebook.
Among those who had, the most commonly used tools blocked access (7 participants) or removed the newsfeed (3 participants).
3 participants currently used such tools; one used \emph{Newsfeed Eradicator} (which removes the newsfeed), another used \emph{Self-control} (which blocks social media), and the third used an ad blocker (which we did not consider a self-control tool).

\hypertarget{overall-facebook-use}{%
\subsection{Overall Facebook use}\label{overall-facebook-use}}

Across all participants and the entire study period, the median number of daily tab visits to Facebook was 23 (min = 5, max = 138).
The median break length between visits to Facebook was 69.5 seconds (min = 11, max = 445).
The median of participants' average amount of daily time spent was approximately 21 minutes (min = 4m, max = 2h 56m).

Often, a number of successive tab visits was logged within a short span of time (e.g., if participants switched back and forth between active application windows).
Following Cheng et al. \citep{chengUnderstandingPerceptionsProblematic2019}, we calculated the number of `sessions' as the number of times where the break between two visits to Facebook was longer than 60 seconds.
The median number of daily sessions on Facebook was 11 (min = 1, max = 101).

\hypertarget{intervention-use-and-perceptions}{%
\subsection{Intervention use and perceptions}\label{intervention-use-and-perceptions}}

The C\textsubscript{goal} extension did not record what participants typed when prompted for their goal, as we wanted to study effects of goal reminders without participants adapting or self-censoring from knowing responses might be read by the researchers.
However, we asked in the interviews how they had used it.
Most said they wrote short, descriptive, but generic notes for what they did (``I would type shorthand in for what I was about to do, so most of the time I would say something like `reply to messages' or just `messages' or `post something on a group' or something like that'', P4).
Some also said they occasionally wrote meaningless or `unsavoury' things when they found the goal prompt annoying or disruptive (``I think sometimes I tried to type in, like, not really proper words and it said, `give me a proper answer' and I was like `dammit'!'', P27).
In C\textsubscript{no-feed}, one participant said the newsfeed occasionally flashed on screen very briefly before being hidden by our script (``sometimes i saw like a millisecond of something and I was like `oh that's interesting, I would like to see that' but then it wasn't there'', P56).

\enlargethispage{4\baselineskip}
In C\textsubscript{control}, a couple of participants said the white background made content stand out less on their screen (``white background definitely makes it harder to\ldots{} I don't think it's easier to read\ldots{}'', P1).
Others, however, found it aesthetically pleasing (``I just liked Facebook more\ldots{} it felt more\ldots{} I mean it felt more Nordic, it wasn't grey and boring, it was white and nice\ldots{}'', P30) and wanted it to persist (``is there a way that I can keep the background white?'', P15).

\hypertarget{rq1-amount-of-use-how-do-goal-reminders-or-removing-the-newsfeed-impact-time-spent-and-visits-made}{%
\subsection{RQ1 (Amount of use): How do goal reminders or removing the newsfeed impact time spent and visits made?}\label{rq1-amount-of-use-how-do-goal-reminders-or-removing-the-newsfeed-impact-time-spent-and-visits-made}}

\begin{figure}
\includegraphics[width=1\linewidth]{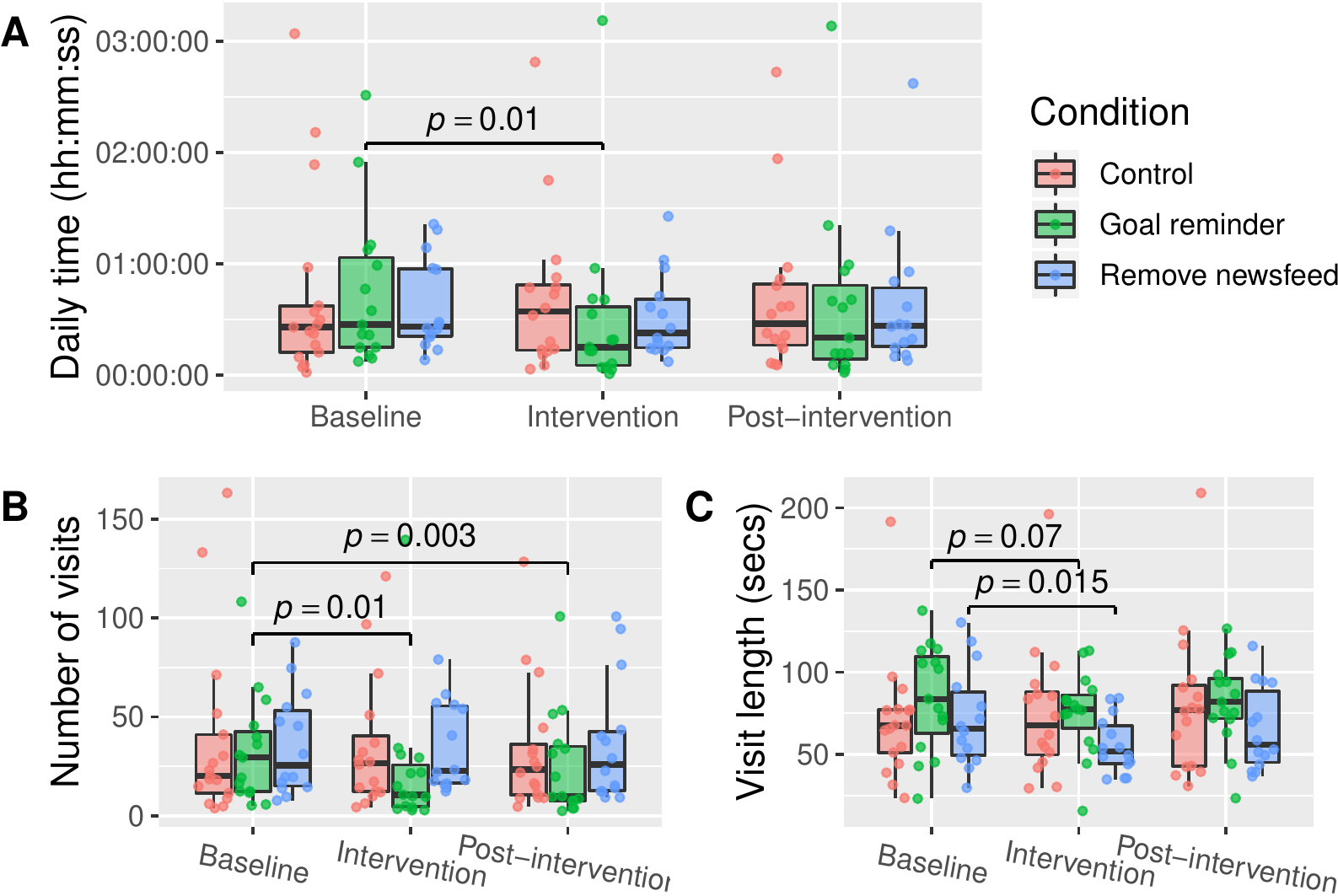} \caption{Time spent and number of visits made to Facebook. Comparing baseline and intervention, goal reminders were associated with less daily time (A), fewer tab visits (B), and a trend towards shorter visits (C). Removing the newsfeed was associated with shorter visits (C). Comparing the post-intervention block to baseline, goal reminders were associated with fewer visits, suggesting an enduring effect of the intervention.}\label{fig:mean-time-boxplot}
\end{figure}

The logging data and qualitative data suggested that C\textsubscript{goal} led to less time spent and fewer and shorter visits, whereas C\textsubscript{no-feed} led to shorter visits (Figure \ref{fig:mean-time-boxplot}):

Usage logging showed that in C\textsubscript{goal}, average daily time on Facebook was significantly lower during the intervention block than in the baseline (median daily time in baseline: 27m 14s,
median in intervention: 15m 5s,
\emph{p} = 0.01, \emph{r} = 0.63, Wilcoxon signed rank test); number of daily visits declined (median number of visits in baseline = 29.4,
median in intervention = 10.6,
\emph{p} = 0.01, \emph{r} = 0.63, Wilcoxon signed rank test);
and there was a trend towards shorter visits (mean tab visit duration in baseline = 1m 25s,
mean in intervention = 1m 15s,
\emph{t}(14) = 1.96, \emph{p} = 0.07, \emph{d} = 0.51).
In C\textsubscript{no-feed}, only visit length declined significantly (mean visit length in baseline = 1m 12s, mean in intervention = 56s, \emph{t}(13) = 2.81, \emph{p} = 0.01, \emph{d} = 0.75).\footnote{Reported effect sizes are Cohen's \emph{d} for t-tests \protect\cite{cohenPowerPrimer1992} and \emph{r} for Wilcoxon signed rank tests \protect\cite{fritzEffectSizeEstimates2012}, computed with the \texttt{rstatix} package for R \protect\cite{kassambaraRstatixPipeFriendlyFramework2019}.}

\enlargethispage{4\baselineskip}
Participants' reports in the surveys and interviews agreed with the logging data:\\
In C\textsubscript{goal}, two common themes were that \textbf{the intervention reduced amount of time on Facebook on laptop} (``yeah i think I used it less and when I was using it I wasn't using it for very long, like a minute maybe'', P45\textsubscript{interview}\footnote{Subscripts indicate whether quotes are from survey free text responses or from post-study interviews, and in some cases also show participants' study condition.}; ``definitely used it a bit less'', P21\textsubscript{interview}) and that \textbf{reduced use was partly caused by the intervention being annoying/stressful} (``This programme made me annoyed thus I would spent {[}sic{]} less time on Facebook'', P32\textsubscript{survey}; ``The changes stressed me to get done with my task and then close facebook'', P40\textsubscript{survey}).

In C\textsubscript{no-feed}, participants had \textbf{mixed opinions on whether or not it reduced amount of use}.
Some felt it reduced their use (``limited overall usage'', P28\textsubscript{survey}, ``I think I used it less erm for shorter periods of time'' P55\textsubscript{interview}) but others felt it only changed their newsfeed use without affecting amount per se (``The lack of newsfeed is welcome \ldots{} Facebook usage on my laptop has not changed/barely changed'', P27\textsubscript{survey}; ``I spent a lot of time actually on facebook but messaging other people and not just looking through my wall'', P54\textsubscript{interview}).

\hypertarget{rq2-patterns-of-use-how-do-goal-reminders-or-removing-the-newsfeed-impact-patterns-of-use}{%
\subsection{RQ2 (Patterns of use): How do goal reminders or removing the newsfeed impact patterns of use?}\label{rq2-patterns-of-use-how-do-goal-reminders-or-removing-the-newsfeed-impact-patterns-of-use}}

\begin{figure}
\includegraphics[width=1\linewidth]{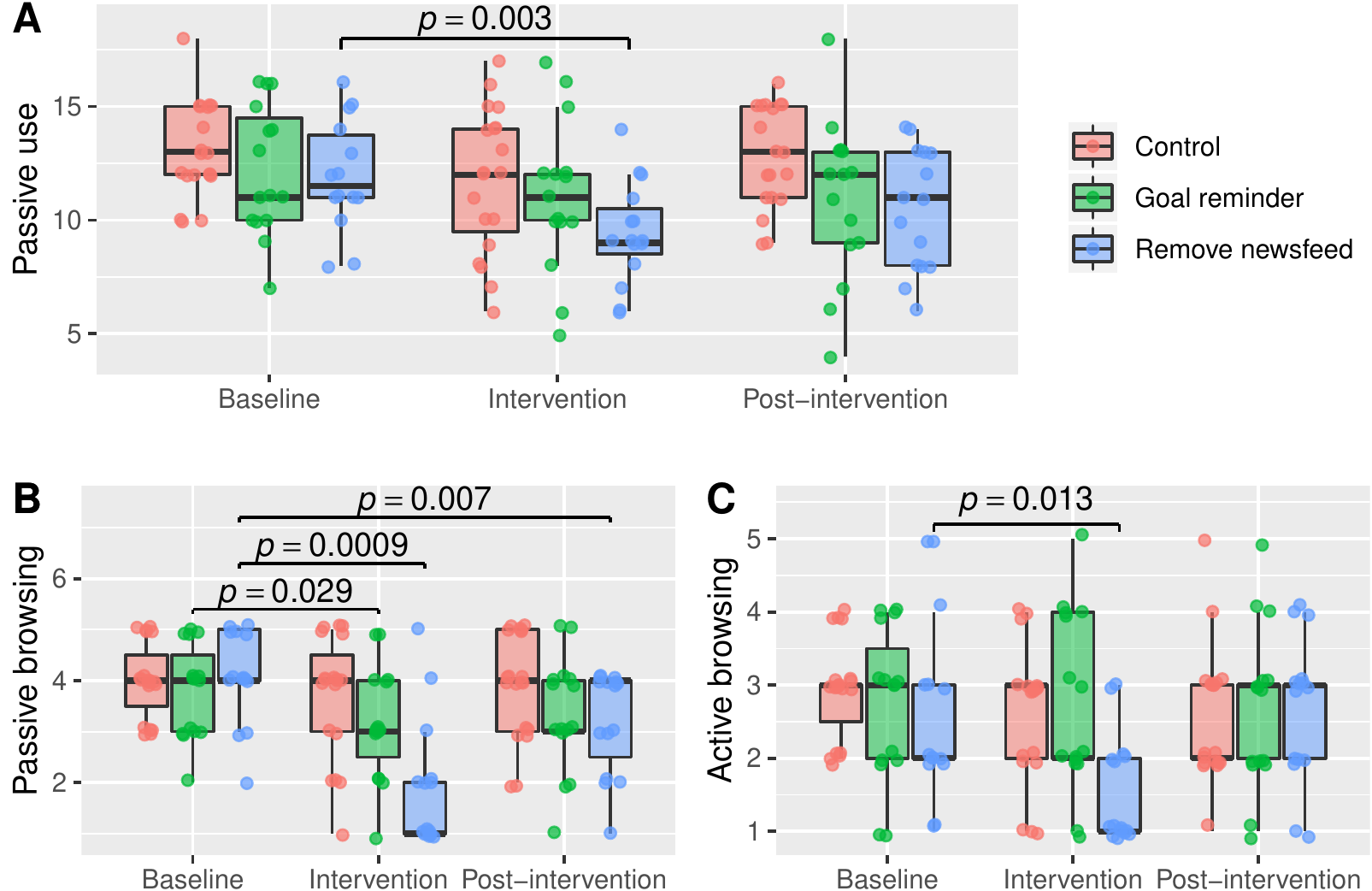} \caption{Scores on the Passive and Active Facebook Use Measure by condition. Comparing the intervention to the baseline block, removing the newsfeed reduced scores on the 'passive' dimension (A), as well as (as expected) individual items 'Browsing the newsfeed passively (without liking or commenting on anything)'s (B) and 'Browsing the newsfeed actively (liking and commenting on posts, pictures and updates)' (C). Goal reminders reduced only passive newsfeed browsing (B). Comparing post-intervention and baseline, removing the newsfeed was associated with reduced passive newsfeed browsing post-intervention (B).}\label{fig:passive-active-plots}
\end{figure}

The logging, survey, and interview data suggested that both C\textsubscript{goal} and C\textsubscript{no-feed} affected patterns of use:
C\textsubscript{goal} selectively reduced passive scrolling of the newsfeed, whereas C\textsubscript{no-feed} (as expected) reduced all behaviour related to the newsfeed (Figure \ref{fig:passive-active-plots}).

Thus, \emph{usage logging} showed that average daily scrolling declined by 42\% in C\textsubscript{goal} (comparing intervention to baseline, \emph{t}(14) = 2.39, \emph{p} = 0.03, \emph{d} = 0.62), and by 73\% in C\textsubscript{no-feed} (\emph{t}(13) = 4.15, \emph{p} = 0.001, \emph{d} = 1.11).
Moreover, in C\textsubscript{no-feed}, the number of times content was liked declined (median number of likes during baseline = 16, median during intervention = 7, \emph{p} = 0.002, \emph{r} = 0.88, Wilcoxon signed rank test).

In the \emph{surveys}, scores on the Passive and Active Facebook Use Measure dimensions showed that participants in C\textsubscript{no-feed} had substantially lower scores on `passive' use in the intervention than in the baseline block (\emph{t}(13) = 4.8, \emph{p} = 0.0003, \emph{d} = 1.28).
We explored effects on more granular elements of Facebook use by comparing baseline and intervention scores separately for each item of the PAUM.\footnote{The reported p-values are not corrected for multiple comparisons --- these should be considered exploratory results to be followed-up with confirmatory studies.}
Two items showed significant variation with condition:
``Browsing the newsfeed passively (without liking or commenting on anything)'' and ``Browsing the newsfeed actively (liking and commenting on posts, pictures and updates)'':
in C\textsubscript{goal}, participants reported less passive, but not active, browsing of the newsfeed during the intervention block compared to baseline (Passive browsing: \emph{p} = 0.03, \emph{r} = 0.57, Active browsing: \emph{p} = 1, \emph{r} = 0.05, Wilcoxon signed rank test).
In C\textsubscript{no-feed}, participants reported less passive as well as less active newsfeed browsing (Passive browsing: \emph{p} = 0.001, \emph{r} = 0.89, Active browsing: \emph{p} = 0.01, \emph{r} = 0.69, Wilcoxon signed rank test).
Moreover, participants in C\textsubscript{no-feed} showed a trend towards lower scores on ``Commenting (on statuses, wall posts, pictures, etc)'' (\emph{p} = 0.09, \emph{r} = 0.46, Wilcoxon signed rank test).

\enlargethispage{1\baselineskip}
The quantitative results were supported by the qualitative data:\\
for participants in both experimental conditions, a recurrent theme was that the interventions caused \textbf{decreased browsing of the newsfeed} (``I did feel very aware when scrolling down my newsfeed, and cut it down'', P19\textsubscript{goal\_survey}; ``definitely meant I spent less time scrolling on newsfeed on my laptop'', P55\textsubscript{no-feed\_survey}), and \textbf{increased use of Facebook for other, more deliberate purposes} (``a big facebook post or whatever not just passively\ldots{}scrolling'', P41\textsubscript{goal\_interview}; ``messaging other people and not just looking through my wall'', P54\textsubscript{no-feed\_interview}).\\
In \textbf{C\textsubscript{goal}}, participants said the effects were driven by the intervention making them \textbf{search for reasons to justify being on the site} (``Being asked why I was opening Facebook was really helpful as it made me question why'', P41\textsubscript{goal\_survey}; ``less likely to aimlessly browse, as I couldn't justify it'', P45\textsubscript{goal\_survey}). In \textbf{C\textsubscript{no-feed}}, participants said the lack of a newsfeed made them \textbf{seek out alternative options that were often more productive and deliberate} (``procrastination was more productive in that I was uhm seeking things out to read or to do that were more intentional, I suppose, and less kind of mindless which I guess the newsfeed is'', P12\textsubscript{no-feed\_interview}). (Changed patterns of use related to perceived control are reported below.)

\hypertarget{rq3-control-how-do-goal-reminders-or-removing-the-newsfeed-impact-perceived-control}{%
\subsection{RQ3 (Control): How do goal reminders or removing the newsfeed impact perceived control?}\label{rq3-control-how-do-goal-reminders-or-removing-the-newsfeed-impact-perceived-control}}

\begin{figure}
\includegraphics[width=1\linewidth]{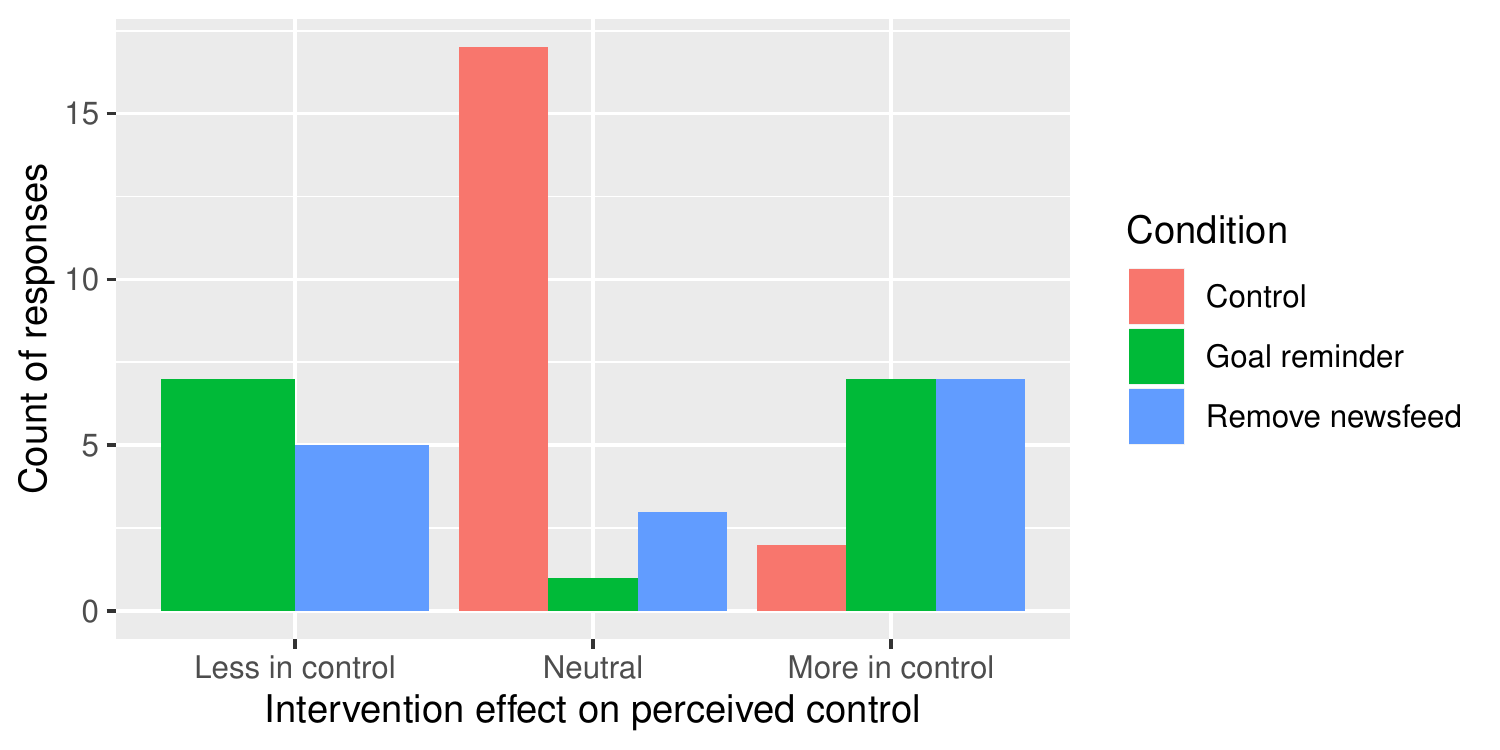} \caption{Responses from survey question on control included in the survey administered by the end of the intervention block: 'During the last two weeks, did the changes made to Facebook on your laptop affect how in control you felt over your use of the site on your laptop?'}\label{fig:control-on-laptop}
\end{figure}

The qualitative data suggested that C\textsubscript{goal} and C\textsubscript{no-feed} supported control in the sense of helping participants avoid unintended use and staying on task, but at the cost of being annoying/frustrating (C\textsubscript{goal}) or leading to fear of missing out (C\textsubscript{no-feed}).

When exploring survey responses in the Multidimensional Facebook Intensity Scale, the only of its four dimensions that showed significant differences between the baseline and intervention blocks was \emph{overuse}:
scores on this measure trended towards decrease during the intervention in all conditions
(C\textsubscript{control}: \emph{t}(18) = 2.37, \emph{p} = 0.03, \emph{d} = 0.54,
C\textsubscript{no-feed}: \emph{t}(13) = 1.99, \emph{p} = 0.07, \emph{d} = 0.53,
C\textsubscript{goal}: t(14) = 1.75, \emph{p} = 0.1, \emph{d} = 0.45),
perhaps suggesting that the study procedure across conditions made participants reflect on use.

\enlargethispage{3\baselineskip}
When asked directly in a survey item following the intervention block whether the changes made to Facebook made them feel less or more in control of their use, C\textsubscript{control} had no impact, while participants seemed divided about the impact of the experimental conditions (Figure \ref{fig:control-on-laptop}).

The qualitative data seemed to provide the explanation:\\
In both C\textsubscript{goal} and C\textsubscript{no-feed}, it was a strong theme in the surveys and interviews that the interventions \textbf{helped participants stay on their intended task during use} (``used it less for stuff that I wasn't intending when I opened it'', P4\textsubscript{goal\_interview}; ``I'll kind of forget that I'm doing work and start scrolling so it was useful to not be able to do that'', P47\textsubscript{no-feed\_interview}).
A subtheme was that this included making it \textbf{easier to disengage from use} (``it's good to get this reminder of `hey you can get off this thing'\,'', P31\textsubscript{goal\_interview}; ``it was easier just to log out, just check what I had to and then leave facebook'' P54\textsubscript{no-feed\_interview}).
In \textbf{C\textsubscript{goal}}, participants said the reason the intervention helped them stay on task was that it \textbf{helped them snap out of automatic use}, that is, stop themselves when they engaged in unintended behaviour (``{[}the reminder{]} sort of snaps you out of that trance, you know what I mean?'', P21\textsubscript{interview}).
In \textbf{C\textsubscript{no-feed}}, participants said the intervention \textbf{stopped unintended behaviours from being triggered} in the first place (``there is nothing here {[}referring to the newsfeed{]}, like `what did I want?', you know, so then I went and contacted the person or looked at the specific thing that I wanted, not what I saw and kinda wanted at the moment'', P56\textsubscript{interview}).

At the same time, there were downsides to the interventions in that \textbf{C\textsubscript{goal} was frequently annoying or frustrating}, especially because it was \textbf{not sensitive to context} (``I use facebook just to message people and I found this extremely annoying because I need to tell someone something and then this thing comes up and I'd just get annoyed\ldots{}'' P32\textsubscript{interview}), and that \textbf{C\textsubscript{no-feed} led to fear of missing out} (``missing out on a lot because actually a lot of the ways I interact with people on facebook is things I see on the newsfeed'', P12\textsubscript{interview}).

\enlargethispage{3\baselineskip}
\hypertarget{cross-device-use}{%
\subsubsection{Cross-device use}\label{cross-device-use}}

When asked if the interventions changed how they used Facebook on smartphone vs laptop, 86\% of participants in C\textsubscript{goal} and 57\% in C\textsubscript{no-feed} answered `Yes' (compared to 6\% in C\textsubscript{control}, \(\chi^2\) = 65.19, \emph{p} \textless{} 0.001).

Unpacking this in the qualitative data, participants in both experimental conditions expressed that \textbf{cross-device access helped them manage the interventions' downsides, while still enjoying the positive effects} (``if I was scrolling through the newsfeed or checking events, then it wouldn't be annoying because I shouldn't be doing that on my laptop while I'm working, and if it was something like sending messages about work, contacting friends and asking for help then I could use my phone'', P40\textsubscript{goal\_interview}; ``I could reap the benefits of the newsfeed but without being sucked into it on two platforms'', P28\textsubscript{no-feed\_survey}), and so they \textbf{sometimes used their smartphone for activities on Facebook the interventions interfered with, but as a more deliberate choice} (``you're working on your laptop, uhm, and then it's very easy to just click new tab, but having to get your phone out\ldots{}'', P19\textsubscript{goal\_interview}; ``the time I did spend on my phone was more, like, focused because I was actually looking for things I missed out on on my laptop'', P55\textsubscript{no-feed\_interview}).

\hypertarget{rq4-post-intervention-effects-do-the-effects-rq1-3-of-goal-reminders-or-removing-the-newsfeed-persist-after-interventions-are-removed}{%
\subsection{RQ4 (Post-intervention effects): Do the effects (RQ1-3) of goal reminders or removing the newsfeed persist after interventions are removed?}\label{rq4-post-intervention-effects-do-the-effects-rq1-3-of-goal-reminders-or-removing-the-newsfeed-persist-after-interventions-are-removed}}

Comparing post-intervention to baseline, C\textsubscript{goal} and C\textsubscript{no-feed} were associated with some persisting effects, with participants in C\textsubscript{goal} engaging in fewer daily visits and some feeling that the intervention helped build a habit of more intentional use, and participants in C\textsubscript{no-feed} engaging in less passive newsfeed browsing.

Thus, in terms of \emph{amount of use}, participants in C\textsubscript{goal} made fewer daily visits post-intervention compared to baseline (median number of daily visits in first baseline = 29.4, median in post-intervention block = 9.8, \emph{p} = 0.003, \emph{r} = 0.72, Wilcoxon signed rank test).\\
In terms of \emph{patterns of use}, participants in C\textsubscript{no-feed} reported less passive browsing of the newsfeed post-intervention compared to baseline (\emph{p} = 0.007, \emph{r} = 0.78, Wilcoxon signed rank test).
In the interviews, some C\textsubscript{no-feed} participants expressed \textbf{feeling less attracted by the newsfeed when it returned} (``I found myself less interested in the newsfeed'', P10\textsubscript{interview}).\\
In terms of \emph{perceived control}, some participants in C\textsubscript{goal} said the intervention helped them build a persisting \textbf{habit of asking themselves what their intention of use was} when visiting the site (``from this week there is a habit being built\ldots{} asking myself why I'm opening Facebook and that habit's perpetuated more or less to this week'', P34\textsubscript{interview}, ``I'm still aware every time I open Facebook, I'm just a bit more aware every time\ldots{} it's not the reflex anymore now that I've had that experience where I have to write everything down'', P1\textsubscript{interview}).

\enlargethispage{3\baselineskip}
\hypertarget{rq5-self-reflection-do-the-interventions-enable-participants-to-reflect-on-their-struggles-in-ways-that-might-inform-the-design-of-more-effective-interventions}{%
\subsection{RQ5 (Self-reflection): Do the interventions enable participants to reflect on their struggles in ways that might inform the design of more effective interventions?}\label{rq5-self-reflection-do-the-interventions-enable-participants-to-reflect-on-their-struggles-in-ways-that-might-inform-the-design-of-more-effective-interventions}}

In the interviews, nearly all participants expressed feeling conflicted about Facebook, in that they found it too useful or engrained in their lives to do without, but also an ongoing source of distraction and self-control struggles.
They readily suggested a range of design solutions to mitigate self-control struggles.
The extent to which interventions were perceived as freely chosen was important to how it was received, and participants did not trust Facebook to provide solutions.

\hypertarget{struggles-with-facebook-use-1}{%
\subsubsection{Struggles with Facebook use}\label{struggles-with-facebook-use-1}}

\textbf{Too useful to do without, but source of distraction and self-control struggles}:
On the one hand, Facebook provided functionality participants could not --- or would not --- do without, especially messaging, events, groups, and pages.
On the other, Facebook was frequently distracting and caused them to waste time and feel frustrated (``I just want\ldots{}to hack myself to have the self-control to, like, not get distracted\ldots{} I literally just use it as distraction'', P42\textsubscript{no-feed}).
In particular, participants struggled to use Facebook in line with their intentions.
Main aspects included
(i) going to the site to do one thing, but then forgetting this goal (``there is one specific trigger that I need to open facebook, but because when I open the page immediately there is tons of information there, like erm notifications, and you scroll down endless streaming\ldots{} so very easily I could be distracted'', P34\textsubscript{goal}),
(ii) internal conflict between short-term gratification and longer-term goals (``might find them {[}videos{]} funny in the short term but when I think about it in the bigger picture it is a complete waste of time'', P48\textsubscript{control}), and
(iii) using Facebook purely out of habit.
In relation to the latter, emotional states, especially boredom, were mentioned as triggers of habitual use (``if I'm in that erm not very motivated state\ldots{} I'll literally just find myself opening it, without even thinking that I'm doing it'', P17\textsubscript{control}).

\hypertarget{specific-suggestions-for-design-solutions}{%
\subsubsection{Specific suggestions for design solutions}\label{specific-suggestions-for-design-solutions}}

Four themes emerged in relation to specific design suggestions for mitigating these struggles:

\textbf{Control over the newsfeed}:
More than half of participants explicitly said the newsfeed did not give them what they wanted, and they desired easy ways to filter it, limit it, or turn it off. Some had tried customising their newsfeeds, but found Facebook's means of doing so tedious and ineffective (``I browse through shit that I don't want to see and I keep on clicking on `I don't like this', `this is not interesting' and of course it keeps on adding new stuff so that doesn't solve the problem basically'', P51\textsubscript{control}).
Solution suggestions included simple ways to filter the newsfeed (``a slider to modify the amount you see people who are on your newsfeed at different percentiles'', P49\textsubscript{goal}, ``two different ones, like you could have a `friends' or like `photos' or something'', P17\textsubscript{control}), reducing the amount of information (``maybe it should be limited to like ten posts and you wouldn't get another ten until the next hour'', P45\textsubscript{goal}, ``if it was instead like blank and then you opt-in to who you actually wanna see on your newsfeed as opposed to opt-out'', P44\textsubscript{no-feed}), or being able to remove it altogether.

\textbf{Raise awareness of time spent or usage goals}:
Participants often lost track of time spent, or of their usage goals, and wanted reminders that raised awareness.
These should be easily accessible (``you wouldn't want it to be buried in settings, something that was actively shown to you I think that would be useful'', P52\textsubscript{control}), and let users judge whether their use was appropriate (``if I saw like `you've spent 2 minutes today', like `great, I've got loads of time that I can waste tomorrow because I've been good today'\,'', P6\textsubscript{goal}).
Participants in C\textsubscript{goal} said the timing and intrusiveness should be calibrated differently to the reminders they experienced in the study (``less in-your-face\ldots{} so maybe more, longer intervals and not the expanding thing\ldots{} if I could change it to longer intervals and maybe a bit less invasive then I think it would actually help'', P4\textsubscript{goal}).

\enlargethispage{3\baselineskip}
\textbf{Remove `addictive' features}:
Participants wished to remove or modify features driving them to use the site.
Specific features mentioned included notifications (``get rid of notifications\ldots{} if I didn't have things popping up every 30 minutes like `this has happened' I don't think I would think about Facebook', P6\textsubscript{goal}), viral videos, and games (``things like game suggestions and like all that sort of stuff I would definitely get rid of cause\ldots{} I don't want to play games \ldots{} `stop bugging me'\,'', P55\textsubscript{no-feed}).
One interesting suggestion was to be able to display content as text-only (``limit it to like text-only posts when you're working so that you're not bothered by videos and algorithms and photos'', P45\textsubscript{goal}).

\textbf{Flexible blocking to meet individual definitions of distraction}:
Participants suggested blocking solutions that could adapt to the type --- or timing --- of use they found distracting.
Thus, some said blocking access altogether was too inflexible to be useful (``there are useful uses of Facebook that aren't just waste of time\ldots{}a blanket, like, `don't do anything on Facebook'\ldots{} it's not practical for those people who have to use Facebook'', P41\textsubscript{goal}).
Suggestions for useful solutions included being able to block or allow specific functionality within Facebook, block access only during specific times (``sync it with a timetable, like lectures or something'', P45\textsubscript{goal}), or even automatically detect if activity is engaged with as a distraction.

\hypertarget{generic-solution-needs}{%
\subsubsection{Generic solution needs}\label{generic-solution-needs}}

\textbf{People differ in what they seek on Facebook and the design solutions they prefer:}
Some participants wanted to block or remove distractions, whereas others preferred less intrusive solutions, such as goal reminders.
Similarly, even though most participants were dissatisfied with the newsfeed, some wanted it to prioritise close ties, whereas others wanted it to prioritise pages they follow (``I wouldn't want to see anyone's posts, I would only want to see posts by things I wanted to follow, whether that's petitions or science papers'', P20\textsubscript{no-feed}).

\textbf{Interventions can `backfire' if overly intrusive and/or not freely chosen:}
Participants felt interventions could make people to rebel against them if too intrusive and/or if they did not feel in charge.
In terms of \emph{intrusiveness}, some felt blocking tools could backfire for this reason (``I feel like most people in their nature, if you have something restrictive\ldots{} then you kinda want to rebel against it'', P56\textsubscript{no-feed}).
In terms of \emph{feeling in control}, some participants suggested this could change their reaction to the very same intervention.
For example, a participant in C\textsubscript{goal} felt the goal reminders were too intrusive and led to resistance (``I got very used to clicking out of it and like, I'm just gonna stay on just out of spite'', P19\textsubscript{goal}), but thought she would react differently if she controlled the reminders herself (``it would be a bit different if it was me, if I could actually write the messages\ldots{} I think that'd help me, and knowing it was me, so it wasn't anyone else'').

\textbf{Scepticism about design solutions coming from Facebook:}
Participants did not trust Facebook to provide effective solutions for mitigating self-control struggles, because this was seen as going against their business interests (''you wonder how much they'd try to just give people the information that doesn't really reflect badly on them'', P36\textsubscript{control}; ``Facebook's interest is for people to spend more time on it 'cause then they'll get more ad revenue, so\ldots{}'', P45\textsubscript{goal}).

\enlargethispage{3\baselineskip}
\hypertarget{discussion}{%
\section{Discussion}\label{discussion}}

\begin{figure}
\includegraphics[width=1\linewidth]{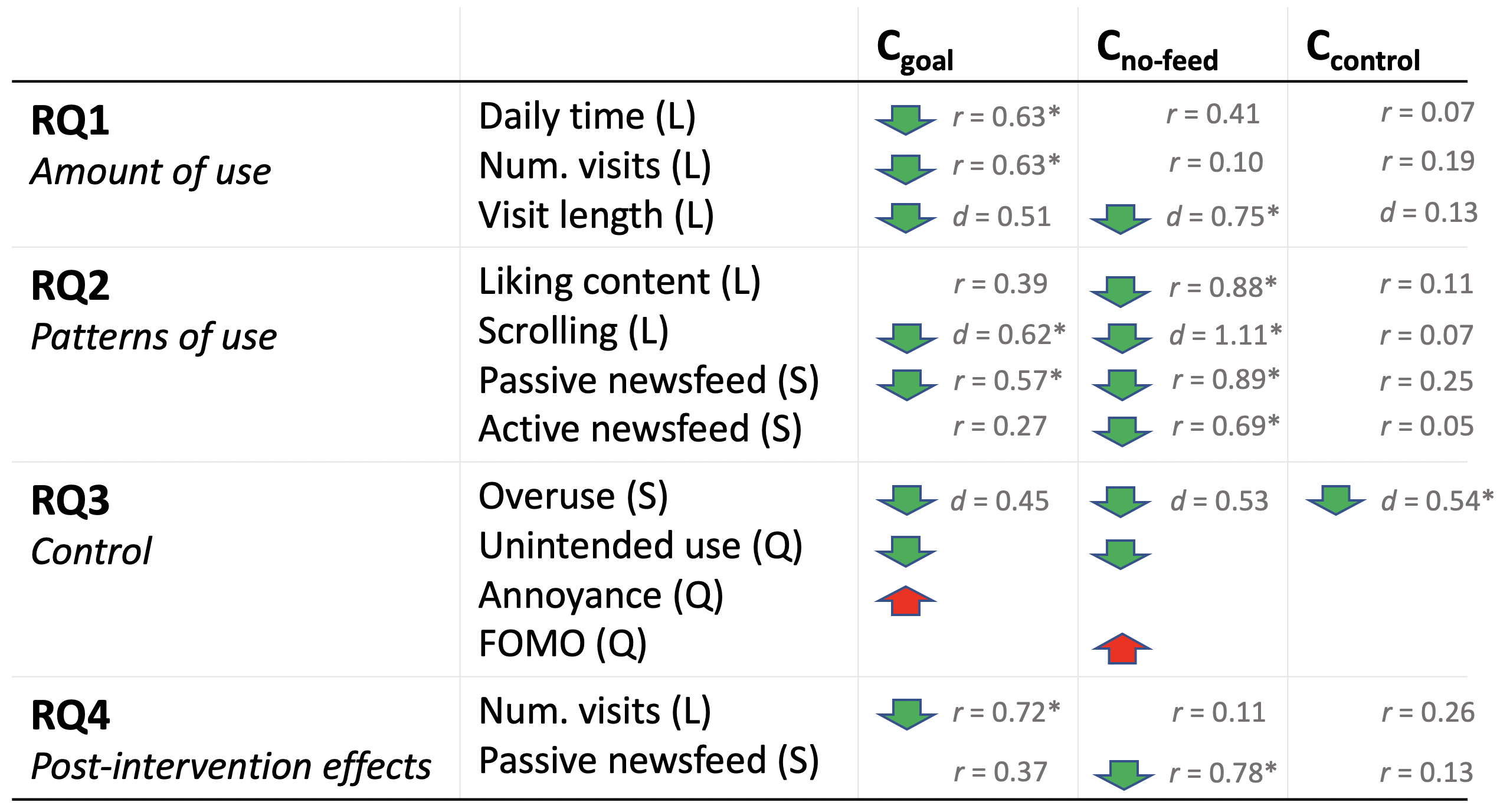} \caption{Summary of main findings for RQ1-4. 'L' = logged usage data, 'S' = quantitative survey data, 'Q' = qualitative data from surveys and interviews. For quantitative data, arrows indicate direction of the effect when \emph{p} \textless{} .10, and effect sizes are marked with an asterisk when \emph{p} \textless{} .05.}\label{fig:results-table}
\end{figure}

Figure \ref{fig:results-table} summarises findings from RQ1-4:
both C\textsubscript{goal} and C\textsubscript{no-feed} reduced unintended Facebook use (RQ3), with the downside that C\textsubscript{goal} was often experienced as annoying and C\textsubscript{no-feed} made some fear missing out on information (cf. ``FOMO'', \citet{przybylskiMotivationalEmotionalBehavioral2013}).
On amount of use (RQ1), C\textsubscript{goal} reduced daily time, number of visits, and visit length, whereas C\textsubscript{no-feed} reduced visit length.
On patterns of use (RQ2), C\textsubscript{goal} and C\textsubscript{no-feed} reduced scrolling and passive newsfeed browsing, and C\textsubscript{no-feed} in addition reduced active newsfeed browsing and amount of content 'liked'.
On post-intervention effects (RQ4), C\textsubscript{goal} was associated with fewer visits and C\textsubscript{no-feed} with less passive newsfeed browsing.\\
In terms of reflections on struggles and solutions (RQ5), participants felt conflicted because Facebook was a source of distraction and self-control struggles but also vital to staying connected, i.e., too useful to avoid.
They suggested specific design solutions related to control over the newsfeed, reminders of time spent and usage goals, removing `addictive' features, and flexible blocking.
Their preferred solutions (as well as the information sought on Facebook) differed, however, and they felt that solutions might `backfire' if overly intrusive and/or not freely chosen.
We now discuss design implications as well as some of the limitations and future work.

\hypertarget{evaluating-the-experimental-interventions}{%
\subsection{Evaluating the experimental interventions}\label{evaluating-the-experimental-interventions}}

Focusing specifically on the ability to use Facebook in line with one's conscious intentions ----- which is at the very core of self-control \citep{duckworthUnpackingSelfControl2015} ----- which of our two experimental interventions is more effective?
Goal reminders and removing the newsfeed represent contrasting, and potentially complementary, strategies.
In our study, both strategies had a positive effect on perceived control and a significant effect on behaviour, with C\textsubscript{goal} helping people 'snap out' of unintended behaviour and C\textsubscript{no-feed} preventing unintended behaviours from being triggered.
While these results suggest that both interventions have potential, as an exploratory study with a restricted sample, further research is needed to draw definitive conclusions about robustness, effect sizes, and individual differences.
However, contextualising our study within related research in psychology and HCI can provide some predictions:

\enlargethispage{3\baselineskip}
One possible approach is to apply a dual systems model of self-regulation (as called for in recent HCI research \citep{coxDesignFrictionsMindful2016, pinderDigitalBehaviourChange2018, adamsMindlessComputingDesigning2015, lyngsSelfControlCyberspaceApplying2019}).
From this perspective, \textbf{goal reminders are a `System 2' intervention} which supports conscious self-control by bringing the goals into working memory that the user wishes to control her behaviour in relation to.
\textbf{Removing the newsfeed is both a `System 1' and `System 2' intervention} which prevents unwanted automatic responses from being triggered by the newsfeed, and supports conscious self-control by preventing attention-grabbing information from crowding out working memory and making the user forget her goal.\\
A recent comprehensive review of digital behaviour change interventions found that providing information about the consequences of behaviour (a System 2 intervention) tends to be unsuccessful, despite being the most common technique.
The authors argued that targeting unconscious habit formation (System 1) should be the focus for interventions that aim at long-term efficacy \citep{pinderDigitalBehaviourChange2018}.
Similarly, psychological research has found that people who are better at self-control tend to develop habits that make their intended behaviour more reliant on automatic processes (System 1) and less on conscious in-the-moment self-control (System 2), and/or reduce their exposure to `temptations' in the first place \citep{gallaMoreResistingTemptation2015, duckworthUnpackingSelfControl2015, duckworthStitchTimeStrategic2016}.
This may be because effective System 2 control depends not only on remembering longer-term goals, but also on one's motivation to exert control relative to those goals, which can fluctuate with emotional state (cf.~participants who said they were more likely to go on Facebook when bored or unmotivated \citep{berkmanSelfControlValueBasedChoice2017, inzlichtWhySelfcontrolSeems2014, Lee2011Mining}).

We therefore expect removing the newsfeed to be more generally effective than goal reminders, because it reduces the amount of potentially distracting information and thus the need for in-the-moment conscious control.
In our study, the qualitative data did suggest that C\textsubscript{goal} fostered a habit of asking oneself about one's purpose when visiting Facebook.
However, given the above, the likelihood of effective control through a habit of goal awareness should depend on what content is available and how that content is perceived:
the more `engaging' the content, the greater the risk that goal awareness will not by itself provide sufficient control motivation \citep{berkmanSelfControlValueBasedChoice2017, ticeEmotionalDistressRegulation2001, Lee2011Mining}.
Goal reminders should therefore exhibit larger variation in effectiveness, and may be less useful for individuals whose newsfeeds contain more attention-grabbing content and/or who struggle more with inhibiting distractions in general. 
This would align with recent findings that those who find Facebook more valuable are also (somewhat paradoxically) more likely to find their use problematic \citep{chengUnderstandingPerceptionsProblematic2019}.
Similarly, prior work suggest that blocking off online distractions is more effective for individual who are more susceptible to social media distractions \citep{markEffectsIndividualDifferences2018} (cf. \citep{Lee2011Mining, miriEmotionRegulation2018}).
These strategies are, however, not mutually exclusive and can be combined in effective interventions, as is already the case in many digital self-control tools (e.g., \emph{Todobook} \citep{yummyappsTodobook2019}, which removes Facebook's newsfeed and replaces it with a to-do list reminding the user of her goals).


\enlargethispage{3\baselineskip}
\hypertarget{designing-future-interventions}{%
\subsection{Designing future interventions}\label{designing-future-interventions}}

Broadly, participants' suggested design solutions related to either \emph{altering the information landscape} (by filtering the newsfeed, removing features driving engagement, or blocking distracting elements) or \emph{raising awareness to help navigation within this landscape} (by adding reminders of time spent or usage goals).
These suggestions could be compared to the many existing interventions on online stores, analysed using a dual systems or other model, and strategies more likely to be effective implemented and evaluated.
Here, we discuss implications of the cross-cutting theme that interventions should be experienced as freely chosen and not overly intrusive to avoid `backfiring' and motivate people to rebel against an intervention instead of being helped by it (cf. \citep{Lee2011Mining}).

Given that participants preferred different interventions --- with some wanting restrictive blocking tools --- it is not a solution to only consider, e.g., non-intrusive addition of user controls \citep{harambamRecommenderSystems2019}.
Rather, designers should keep in mind that the effectiveness of the exact same restriction or intrusion may depend on whether it is perceived by the user as self-imposed or externally imposed \citep{brookEcologicalFootprintFeedback2011, Swim2013, Bryan2010}.
An implication is that interventions should be carefully framed as being supportive of the user's personal goals (cf. \citep{Swim2013, Bandura1982}).
For example, blocking tools may wish to remind the user why their past self decided to impose restrictions on their present self \citep{duckworthWillpowerStrategiesReducing2018}.
Current examples `in the wild' include browser extensions for website blocking that display motivational quotes or task reminders when users navigate to distracting sites \citep{lyngsSelfControlCyberspaceApplying2019}.

One exciting avenue for future tools is systems that can learn the user's personal definition of distraction and in what contexts to, e.g., automatically impose or not impose limits.
This was suggested by one of our participants, and is being explored in some HCI research, e.g. \href{https://habitlab.github.io}{\emph{HabitLab}}, which rotates between interventions to discover what best helps a user limit time on specific websites \citep{Kovacs2018}.
A useful such system in the context of Facebook would not simply limit time, but rather assist the user in carrying out their goals, for example by dynamically blocking elements such as the newsfeed if the user's current goal is to create an event.
Such a hypothetical system could be highly useful, but it would be crucial to its success that its interventions were perceived by the user as being in her own interest.
In addition, it would need to \emph{really} understand the user to be functional \citep{lyngsTellMeWhat2018}, creating a possible trade-off between privacy and the `fit' of the intervention.
Facebook itself, with its deep knowledge of user behaviour, might be in the best position to explore this approach, but we note that participants in our study were deeply sceptical about Facebook's motivations and did not expect design solutions coming from Facebook to be `on their side' (cf. \citep{Creswick2019, PerezVallejos2017}).

\hypertarget{limitations-and-future-work}{%
\subsection{Limitations and future work}\label{limitations-and-future-work}}

\textbf{Confounding variables:}
A possible criticism is that less scrolling and shorter visits when removing the newsfeed occur simply because there is nothing to scroll.
We note that removing the newsfeed did not make scrolling impossible --- it remained relevant on all other pages than the home screen --- and thus scrolling remains a useful measure.
Moreover, reduced time is often an explicit goal for users, and so time spent in the face of reduced content is a relevant outcome.\\
\textbf{Lack of cross-device tracking:}
We logged Facebook use on laptop only and did not quantify effects of the interventions on cross-device use.
It is important in future work to assess potential `spillover' effects between devices when applying interventions meant to scaffold self-control \citep{kovacsConservationProcrastinationProductivity2019, lascauWhyAreCrossDevice2019, kimTechnologySupportedBehavior2017}, and so we encourage follow-up studies to explore how our methods could be supplemented by, e.g., smartphone logging.\\
\textbf{Retrospective self-report:}
In the surveys and interviews, participants retrospectively reported their experience, which is subject to recall biases \citep{Kahneman2005, redelmeierPatientsMemoriesPainful1996}.
As self-control often involves one's past self setting goals for one's future self (e.g., in blocking tools), retrospective reflection is very informative \citep{lyngsTellMeWhat2018}, but it would be interesting in future research to include experience sampling methods to assess in-the-moment experience \citep{reineckeSlackingWindingExperience2016}.\\
\textbf{Granular interventions and usage measures:}
Standard measures of Facebook use were not optimal for assessing granular interventions on laptop only:
most measures consider global use, and factor into broad dimensions.
For example, we found the Passive and Active Facebook Use Measure's overall dimensions too broad to capture the behavioural changes our interventions introduced.
We flag this as a consideration for future study designs.\\
\textbf{Sampling:}
Our sample size was restricted, to allow us to conduct interviews with all participants, and further research is required to assess whether our exploratory results will replicate (ideally in pre-registered studies with minimum sample size guided by our effect size estimates, cf. \citep{cockburnHARKNoMore2018}).
Moreover, our recruitment was restricted to university students.
Whereas previous research suggests that struggles with Facebook use are particularly pronounced in this population, and that finding effective interventions in this population therefore is important, further research is needed to assess how our findings might generalise.
Finally, our recruitment process may have selected for participants who were highly motivated to change their use of Facebook and/or who used it extensively.
Motivation is central to self-control \cite{inzlichtEmotionalFoundationsCognitive2015}, but we did not assess this explicitly.
Our participants' baseline levels of logged use, and scores on the Multidimensional Facebook Intensity Scale, were fairly average compared to previous studies (\citep{phuFacebookUseIts2019, wangContextCollegeStudents2018}, see supplementary analysis on \href{https://osf.io/qtg7h/}{osf.io/qtg7h}), future work will benefit from explicitly measuring participants' level of motivation.

\enlargethispage{3\baselineskip}
\hypertarget{conclusion}{%
\section{Conclusion}\label{conclusion}}

Imagining what success for digital self-control on Facebook and beyond looks like is not an academic exercise, but a practical and urgent concern as evidenced by the recent hearing on `Persuasive Technology' in the US senate \citep{OptimizingEngagementUnderstanding2019}, and a UK All Party Parliamentary Group's call for a `duty of care' to be established on social media companies \citep{allpartyparliamentarygrouponsocialmediaandyoungpeoplesmentalhealthandwellbeingNewFiltersManageImpact2019}.
We encourage future HCI work in this space to assess possible design interventions with open and transparent research methods, to provide the evidence base needed to assist regulators in moving towards a benevolent future \citep{GrimpeRRIHCI2014}.

\hypertarget{acknowledgements}{%
\section{Acknowledgements}\label{acknowledgements}}

We thank Felix Epp for assistance with the ROSE extension;
Michael Inzlicht and Nick Yeung for feedback on the study design;
and Nadia Flensted H\o gholt for feedback on the study design, participant communication, and design of Facebook usage visualisations.

\balance{}

\bibliographystyle{SIGCHI-Reference-Format}
\bibliography{references.bib}


\begin{thebibliography}{00}


\ifx \showCODEN    \undefined \def \showCODEN     #1{\unskip}     \fi
\ifx \showDOI      \undefined \def \showDOI       #1{{\tt DOI:}\penalty0{#1}\ }
  \fi
\ifx \showISBNx    \undefined \def \showISBNx     #1{\unskip}     \fi
\ifx \showISBNxiii \undefined \def \showISBNxiii  #1{\unskip}     \fi
\ifx \showISSN     \undefined \def \showISSN      #1{\unskip}     \fi
\ifx \showLCCN     \undefined \def \showLCCN      #1{\unskip}     \fi
\ifx \shownote     \undefined \def \shownote      #1{#1}          \fi
\ifx \showarticletitle \undefined \def \showarticletitle #1{#1}   \fi
\ifx \showURL      \undefined \def \showURL       #1{#1}          \fi

\bibitem{adamsMindlessComputingDesigning2015}
{Alexander~T. Adams}, {Jean Costa}, {Malte~F. Jung}, {and} {Tanzeem Choudhury}.
  2015.
\newblock \showarticletitle{Mindless {{Computing}}: {{Designing Technologies}}
  to {{Subtly Influence Behavior}}}. In {\em Proceedings of the 2015 {{ACM
  International Joint Conference}} on {{Pervasive}} and {{Ubiquitous
  Computing}}}. {ACM}, 719--730.
\newblock
\showDOI{%
\url{http://dx.doi.org/10.1145/2750858.2805843}}


\bibitem{al-dubaiAdverseHealthEffects2013}
{Sami Abdo~Radman Al-Dubai}, {Kurubaran Ganasegeran}, {Mustafa Ahmed~Mahdi
  Al-Shagga}, {Hematram Yadav}, {and} {John~T. Arokiasamy}. 2013.
\newblock Adverse {{Health Effects}} and {{Unhealthy Behaviors}} among
  {{Medical Students Using Facebook}}.
\newblock https://www.hindawi.com/journals/tswj/2013/465161/.   (2013).
\newblock
\showDOI{%
\url{http://dx.doi.org/10.1155/2013/465161}}


\bibitem{allpartyparliamentarygrouponsocialmediaandyoungpeoplesmentalhealthandwellbeingNewFiltersManageImpact2019}
{{All Party Parliamentary Group on Social Media and Young People's Mental
  Health and Wellbeing}}. 2019.
\newblock {\em \#{{NewFilters}} to Manage the Impact of Social Media on Young
  People's Mental Health and Wellbeing}.
\newblock {T}echnical {R}eport. {UK Parliament}.
\newblock


\bibitem{allcottWelfareEffectsSocial2019}
{Hunt Allcott}, {Luca Braghieri}, {Sarah Eichmeyer}, {and} {Matthew Gentzkow}.
  2019.
\newblock {\em The {{Welfare Effects}} of {{Social Media}}}.
\newblock Working {{Paper}} 25514. {National Bureau of Economic Research}.
\newblock
\showDOI{%
\url{http://dx.doi.org/10.3386/w25514}}


\bibitem{andreassenDevelopmentFacebookAddiction2012}
{Cecilie~Schou Andreassen}, {Torbj{\o}rn Torsheim}, {Geir~Scott Brunborg},
  {and} {Staale Pallesen}. 2012.
\newblock \showarticletitle{Development of a {{Facebook Addiction Scale}}}.
\newblock {\em Psychological Reports\/} {110}, 2 (apr 2012), 501--517.
\newblock
\showISSN{0033-2941}
\showDOI{%
\url{http://dx.doi.org/10.2466/02.09.18.PR0.110.2.501-517}}


\bibitem{yummyappsTodobook2019}
{Yummy Apps}. 2019.
\newblock Todobook.
\newblock   (May 2019).
\newblock


\bibitem{Bandura1982}
{Albert Bandura}. 1982.
\newblock \showarticletitle{Self-efficacy mechanism in human agency.}
\newblock {\em American Psychologist\/} {37}, 2 (1982), 122--147.
\newblock
\showDOI{%
\url{http://dx.doi.org/10.1037/0003-066x.37.2.122}}


\bibitem{banyaiProblematicSocialMedia2017}
{Fanni B{\a'a}nyai}, {{\a'A}gnes Zsila}, {Orsolya Kir{\a'a}ly}, {Aniko Maraz},
  {Zsuzsanna Elekes}, {Mark~D. Griffiths}, {Cecilie~Schou Andreassen}, {and}
  {Zsolt Demetrovics}. 09-Jan-2017.
\newblock \showarticletitle{Problematic {{Social Media Use}}: {{Results}} from
  a {{Large}}-{{Scale Nationally Representative Adolescent Sample}}}.
\newblock {\em PLOS ONE\/} {12}, 1 (09-Jan-2017), e0169839.
\newblock
\showISSN{1932-6203}
\showDOI{%
\url{http://dx.doi.org/10.1371/journal.pone.0169839}}


\bibitem{berkmanSelfControlValueBasedChoice2017}
{Elliot~T Berkman}, {Cendri~A Hutcherson}, {Jordan~L Livingston}, {Lauren~E
  Kahn}, {and} {Michael Inzlicht}. 2017.
\newblock \showarticletitle{Self-{{Control}} as {{Value}}-{{Based Choice}}}.
\newblock {\em Current Directions in Psychological Science\/} {26}, 5 (2017),
  422--428.
\newblock
\showDOI{%
\url{http://dx.doi.org/10.1177/0963721417704394}}


\bibitem{bootPervasiveProblemPlacebos2013}
{Walter~R. Boot}, {Daniel~J. Simons}, {Cary Stothart}, {and} {Cassie Stutts}.
  2013.
\newblock \showarticletitle{The Pervasive Problem with Placebos in Psychology}.
\newblock {\em Perspectives on Psychological Science\/} {8}, 4 (jul 2013),
  445--454.
\newblock
\showDOI{%
\url{http://dx.doi.org/10.1177/1745691613491271}}


\bibitem{brookEcologicalFootprintFeedback2011}
{Amara Brook}. 2011.
\newblock \showarticletitle{Ecological Footprint Feedback: {{Motivating}} or
  Discouraging?}
\newblock {\em Social Influence\/} {6}, 2 (April 2011), 113--128.
\newblock
\showISSN{1553-4510}
\showDOI{%
\url{http://dx.doi.org/10.1080/15534510.2011.566801}}


\bibitem{Bryan2010}
{Gharad Bryan}, {Dean Karlan}, {and} {Scott Nelson}. 2010.
\newblock \showarticletitle{Commitment Devices}.
\newblock {\em Annual Review of Economics\/} {2}, 1 (Sept. 2010), 671--698.
\newblock
\showDOI{%
\url{http://dx.doi.org/10.1146/annurev.economics.102308.124324}}


\bibitem{burkeRelationshipFacebookUse2016}
{Moira Burke} {and} {Robert~E. Kraut}. 2016.
\newblock \showarticletitle{The {{Relationship Between Facebook Use}} and
  {{Well}}-{{Being Depends}} on {{Communication Type}} and {{Tie Strength}}}.
\newblock {\em Journal of Computer-Mediated Communication\/} {21}, 4 (2016),
  265--281.
\newblock
\showISSN{1083-6101}
\showDOI{%
\url{http://dx.doi.org/10.1111/jcc4.12162}}


\bibitem{burkeSocialNetworkActivity2010}
{Moira Burke}, {Cameron Marlow}, {and} {Thomas Lento}. 2010.
\newblock \showarticletitle{Social {{Network Activity}} and {{Social
  Well}}-Being}. In {\em Proceedings of the {{SIGCHI Conference}} on {{Human
  Factors}} in {{Computing Systems}}} {\em ({{CHI}} '10)}. {ACM}, New York, NY,
  USA, 1909--1912.
\newblock
\showISBNx{978-1-60558-929-9}
\showDOI{%
\url{http://dx.doi.org/10.1145/1753326.1753613}}


\bibitem{chenSharingLikingCommenting2013}
{Wenhong Chen} {and} {Kye-Hyoung Lee}. 2013.
\newblock \showarticletitle{Sharing, Liking, Commenting, and Distressed?
  {{The}} Pathway between {{Facebook}} Interaction and Psychological Distress}.
\newblock {\em Cyberpsychology, Behavior and Social Networking\/} {16}, 10 (oct
  2013), 728--734.
\newblock
\showISSN{2152-2723}
\showDOI{%
\url{http://dx.doi.org/10.1089/cyber.2012.0272}}


\bibitem{chengUnderstandingPerceptionsProblematic2019}
{Justin Cheng}, {Moira Burke}, {and} {Elena~Goetz Davis}. 2019.
\newblock \showarticletitle{Understanding {{Perceptions}} of {{Problematic
  Facebook Use}}: {{When People Experience Negative Life Impact}} and a
  {{Lack}} of {{Control}}}. In {\em Proceedings of the 2019 {{CHI Conference}}
  on {{Human Factors}} in {{Computing Systems}}} {\em ({{CHI}} '19)}. {ACM},
  New York, NY, USA, 199:1--199:13.
\newblock
\showISBNx{978-1-4503-5970-2}
\showDOI{%
\url{http://dx.doi.org/10.1145/3290605.3300429}}


\bibitem{cockburnHARKNoMore2018}
{Andy Cockburn}, {Carl Gutwin}, {and} {Alan Dix}. 2018.
\newblock \showarticletitle{{{HARK No More}}: {{On}} the {{Preregistration}} of
  {{CHI Experiments}}}. In {\em Proceedings of the 2018 {{CHI Conference}} on
  {{Human Factors}} in {{Computing Systems}}} {\em ({{CHI}} '18)}. {ACM}, {New
  York, NY, USA}, 141:1--141:12.
\newblock
\showISBNx{978-1-4503-5620-6}
\showDOI{%
\url{http://dx.doi.org/10.1145/3173574.3173715}}


\bibitem{cohenPowerPrimer1992}
{Jacob Cohen}. 1992.
\newblock \showarticletitle{A Power Primer}.
\newblock {\em Psychological Bulletin\/} {112}, 1 (1992), 155--159.
\newblock
\showISSN{1939-1455(Electronic),0033-2909(Print)}
\showDOI{%
\url{http://dx.doi.org/10.1037/0033-2909.112.1.155}}


\bibitem{coxDesignFrictionsMindful2016}
{Anna~L Cox}, {Sandy J~J Gould}, {Marta~E Cecchinato}, {Ioanna Iacovides},
  {and} {Ian Renfree}. 2016.
\newblock \showarticletitle{Design {{Frictions}} for {{Mindful Interactions}}:
  {{The Case}} for {{Microboundaries}}}. In {\em Proceedings of the 2016 {{CHI
  Conference Extended Abstracts}} on {{Human Factors}} in {{Computing
  Systems}}} {\em ({{CHI EA}} '16)}. {ACM}, {New York, NY, USA}, 1389--1397.
\newblock
\showISBNx{978-1-4503-4082-3}
\showDOI{%
\url{http://dx.doi.org/10.1145/2851581.2892410}}


\bibitem{Creswick2019}
{Helen Creswick}, {Liz Dowthwaite}, {Ansgar Koene}, {Elvira~Perez Vallejos},
  {Virginia Portillo}, {Monica Cano}, {and} {Christopher Woodard}. 2019.
\newblock \showarticletitle{{\textquotedblleft}{\ldots} They don't really
  listen to people{\textquotedblright}}.
\newblock {\em Journal of Information, Communication and Ethics in Society\/}
  {17}, 2 (May 2019), 167--182.
\newblock
\showDOI{%
\url{http://dx.doi.org/10.1108/jices-11-2018-0090}}


\bibitem{duckworthWillpowerStrategiesReducing2018}
{Angela~L. Duckworth}, {Katherine~L. Milkman}, {and} {David Laibson}. 2018.
\newblock \showarticletitle{Beyond {{Willpower}}: {{Strategies}} for {{Reducing
  Failures}} of {{Self}}-{{Control}}}.
\newblock {\em Psychological Science in the Public Interest\/} {19}, 3 (Dec.
  2018), 102--129.
\newblock
\showISSN{1529-1006}
\showDOI{%
\url{http://dx.doi.org/10.1177/1529100618821893}}


\bibitem{duckworthUnpackingSelfControl2015}
{Angela~L. Duckworth} {and} {Laurence Steinberg}. 2015.
\newblock \showarticletitle{Unpacking {{Self}}-{{Control}}.}
\newblock {\em Child development perspectives\/} {9}, 1 (2015), 32--37.
\newblock
\showISSN{1750-8592}
\showDOI{%
\url{http://dx.doi.org/10.1111/cdep.12107}}


\bibitem{duckworthStitchTimeStrategic2016}
{Angela~L. Duckworth}, {Rachel~E. White}, {Alyssa~J. Matteucci}, {Annie
  Shearer}, {and} {James~J. Gross}. 2016.
\newblock \showarticletitle{A {{Stitch}} in {{Time}}: {{Strategic
  Self}}-{{Control}} in {{High School}} and {{College Students}}}.
\newblock {\em Journal of Educational Psychology\/} {108}, 3 (apr 2016),
  329--341.
\newblock
\showISSN{0022-0663}
\showDOI{%
\url{http://dx.doi.org/10.1037/edu0000062}}


\bibitem{ellisSmartphoneUsageScales2019}
{David~A. Ellis}, {Brittany~I. Davidson}, {Heather Shaw}, {and} {Kristoffer
  Geyer}. 2019.
\newblock \showarticletitle{Do Smartphone Usage Scales Predict Behavior?}
\newblock {\em International Journal of Human-Computer Studies\/} (may 2019).
\newblock
\showISSN{1071-5819}
\showDOI{%
\url{http://dx.doi.org/10.1016/j.ijhcs.2019.05.004}}


\bibitem{ellisDigitalTracesBehaviour2018}
{David~A. Ellis}, {Linda~K. Kaye}, {Thomas~D.W. Wilcockson}, {and}
  {Francesca~C. Ryding}. 2018.
\newblock \showarticletitle{Digital {{Traces}} of {{Behaviour Within
  Addiction}}: {{Response}} to {{Griffiths}} (2017)}.
\newblock {\em International Journal of Mental Health and Addiction\/} {16}, 1
  (feb 2018), 240--245.
\newblock
\showISSN{1557-1882}
\showDOI{%
\url{http://dx.doi.org/10.1007/s11469-017-9855-7}}


\bibitem{ellisonBenefitsFacebookFriends2007}
{Nicole~B. Ellison}, {Charles Steinfield}, {and} {Cliff Lampe}. 2007.
\newblock \showarticletitle{The {{Benefits}} of {{Facebook}} ``{{Friends}}:''
  {{Social Capital}} and {{College Students}}' {{Use}} of {{Online Social
  Network Sites}}}.
\newblock {\em Journal of Computer-Mediated Communication\/} {12}, 4 (jul
  2007), 1143--1168.
\newblock
\showDOI{%
\url{http://dx.doi.org/10.1111/j.1083-6101.2007.00367.x}}


\bibitem{forstyonahFocusbook2016}
{Yonah Forst}. 2016.
\newblock Focusbook.
\newblock   (2016).
\newblock


\bibitem{fritzEffectSizeEstimates2012}
{Catherine~O. Fritz}, {Peter~E. Morris}, {and} {Jennifer~J. Richler}. 2012.
\newblock \showarticletitle{Effect Size Estimates: Current Use, Calculations,
  and Interpretation}.
\newblock {\em Journal of Experimental Psychology. General\/} {141}, 1 (feb
  2012), 2--18.
\newblock
\showISSN{1939-2222}
\showDOI{%
\url{http://dx.doi.org/10.1037/a0024338}}


\bibitem{gallaMoreResistingTemptation2015}
{Brian~M. Galla} {and} {Angela~L. Duckworth}. 2015.
\newblock \showarticletitle{More {{Than Resisting Temptation}}: {{Beneficial
  Habits Mediate}} the {{Relationship Between Self}}-{{Control}} and {{Positive
  Life Outcomes}}}.
\newblock {\em Journal of Personality and Social Psychology\/} {109}, 3 (2015),
  No Pagination Specified.
\newblock
\showISSN{1939-1315(Electronic);0022-3514(Print)}
\showDOI{%
\url{http://dx.doi.org/10.1037/pspp0000026}}


\bibitem{gersonPassiveActiveFacebook2017}
{Jennifer Gerson}, {Anke~C. Plagnol}, {and} {Philip~J. Corr}. 2017.
\newblock \showarticletitle{Passive and {{Active Facebook Use Measure}}
  ({{PAUM}}): {{Validation}} and Relationship to the {{Reinforcement
  Sensitivity Theory}}}.
\newblock {\em Personality and Individual Differences\/}  {117} (oct 2017),
  81--90.
\newblock
\showISSN{0191-8869}
\showDOI{%
\url{http://dx.doi.org/10.1016/j.paid.2017.05.034}}


\bibitem{goslingVeryBriefMeasure2003}
{Samuel~D. Gosling}, {Peter~J. Rentfrow}, {and} {William~B. {Swann Jr.}} 2003.
\newblock \showarticletitle{A Very Brief Measure of the {{Big}}-{{Five}}
  Personality Domains}.
\newblock {\em Journal of Research in Personality\/} {37}, 6 (2003), 504--528.
\newblock
\showISSN{0092-6566(Print)}
\showDOI{%
\url{http://dx.doi.org/10.1016/S0092-6566(03)00046-1}}


\bibitem{grieveFacetofaceFacebookCan2013}
{Rachel Grieve}, {Michaelle Indian}, {Kate Witteveen}, {G. {Anne Tolan}}, {and}
  {Jessica Marrington}. 2013.
\newblock \showarticletitle{Face-to-Face or {{Facebook}}: {{Can}} Social
  Connectedness Be Derived Online?}
\newblock {\em Computers in Human Behavior\/} {29}, 3 (may 2013), 604--609.
\newblock
\showISSN{0747-5632}
\showDOI{%
\url{http://dx.doi.org/10.1016/j.chb.2012.11.017}}


\bibitem{GrimpeRRIHCI2014}
{Barbara Grimpe}, {Mark Hartswood}, {and} {Marina Jirotka}. 2014.
\newblock \showarticletitle{Towards a Closer Dialogue Between Policy and
  Practice: Responsible Design in HCI}. In {\em Proceedings of the SIGCHI
  Conference on Human Factors in Computing Systems} {\em (CHI '14)}. ACM, New
  York, NY, USA, 2965--2974.
\newblock
\showISBNx{978-1-4503-2473-1}
\showDOI{%
\url{http://dx.doi.org/10.1145/2556288.2557364}}


\bibitem{grosserFacebookDemetricator2019}
{Ben Grosser}. 2019.
\newblock Facebook {{Demetricator}}.
\newblock   (jul 2019).
\newblock


\bibitem{guedesInternetAddictionExcessive2016}
{Eduardo Guedes}, {Federica Sancassiani}, {Mauro~Giovani Carta}, {Carlos
  Campos}, {Sergio Machado}, {Anna Lucia~Spear King}, {and} {Antonio~Egidio
  Nardi}. 2016.
\newblock \showarticletitle{Internet {{Addiction}} and {{Excessive Social
  Networks Use}}: {{What About Facebook}}?}
\newblock {\em Clinical Practice and Epidemiology in Mental Health : CP \&
  EMH\/}  {12} (jun 2016), 43--48.
\newblock
\showISSN{1745-0179}
\showDOI{%
\url{http://dx.doi.org/10.2174/1745017901612010043}}


\bibitem{guptaInclassDistractionsRole2016}
{Natasha Gupta} {and} {Julia~D. Irwin}. 2016.
\newblock \showarticletitle{In-Class Distractions: {{The}} Role of {{Facebook}}
  and the Primary Learning Task}.
\newblock {\em Computers in Human Behavior\/}  {55} (feb 2016), 1165--1178.
\newblock
\showISSN{0747-5632}
\showDOI{%
\url{http://dx.doi.org/10.1016/j.chb.2014.10.022}}


\bibitem{harambamRecommenderSystems2019}
{Jaron Harambam}, {Dimitrios Bountouridis}, {Mykola Makhortykh}, {and} {Joris
  van Hoboken}. 2019.
\newblock \showarticletitle{Designing for the Better by Taking Users into
  Account: A Qualitative Evaluation of User Control Mechanisms in (News)
  Recommender Systems}. In {\em Proceedings of the 13th ACM Conference on
  Recommender Systems} {\em (RecSys '19)}. ACM, New York, NY, USA, 69--77.
\newblock
\showISBNx{978-1-4503-6243-6}
\showDOI{%
\url{http://dx.doi.org/10.1145/3298689.3347014}}


\bibitem{hinikerMyTimeDesigningEvaluating2016}
{Alexis Hiniker}, {Sungsoo~Ray Hong}, {Tadayoshi Kohno}, {and} {Julie~A
  Kientz}. 2016.
\newblock \showarticletitle{{{MyTime}}: {{Designing}} and {{Evaluating}} an
  {{Intervention}} for {{Smartphone Non}}-{{Use}}}.
\newblock {\em Proceedings of the 2016 CHI Conference on Human Factors in
  Computing Systems\/} (2016), 4746--4757.
\newblock
\showDOI{%
\url{http://dx.doi.org/10.1145/2858036.2858403}}


\bibitem{inzlichtEmotionalFoundationsCognitive2015}
{Michael Inzlicht}, {Bruce~D. Bartholow}, {and} {Jacob~B. Hirsh}. 2015.
\newblock \showarticletitle{Emotional Foundations of Cognitive Control}.
\newblock {\em Trends in Cognitive Sciences\/} {19}, 3 (March 2015), 126--132.
\newblock
\showDOI{%
\url{http://dx.doi.org/10.1016/j.tics.2015.01.004}}


\bibitem{inzlichtWhySelfcontrolSeems2014}
{Michael Inzlicht}, {Brandon~J. Schmeichel}, {and} {C.~Neil Macrae}. 2014.
\newblock \showarticletitle{Why Self-Control Seems (but May Not Be) Limited}.
\newblock {\em Trends in Cognitive Sciences\/} {18}, 3 (2014), 127--133.
\newblock
\showISSN{13646613}
\showDOI{%
\url{http://dx.doi.org/10.1016/j.tics.2013.12.009}}


\bibitem{jafarkarimiFacebookAddictionMalaysian2016}
{Hosein Jafarkarimi}, {Alex Tze~Hiang Sim}, {Robab Saadatdoost}, {and} {Jee~Mei
  Hee}. 2016.
\newblock \showarticletitle{Facebook {{Addiction}} among {{Malaysian
  Students}}}.
\newblock {\em International Journal of Information and Education Technology\/}
  {6}, 6 (2016), 465--469.
\newblock
\showISSN{20103689}
\showDOI{%
\url{http://dx.doi.org/10.7763/IJIET.2016.V6.733}}


\bibitem{jdevNewsFeedEradicator2019}
{{JDev}}. 2019.
\newblock News {{Feed Eradicator}} for {{Facebook}}.
\newblock   (jul 2019).
\newblock


\bibitem{Kahneman2005}
{Daniel Kahneman} {and} {Jason Riis}. 2005.
\newblock \showarticletitle{Living, and Thinking about It: {{Two}} Perspectives
  on Life}.
\newblock In {\em The {{Science}} of {{Well}}-{{Being}}}, {F.~A. Huppert},
  {N.~Baylis}, {and} {B.~Keverne} (Eds.). {Oxford University Press}, {New
  York}, 285--304.
\newblock
\showISBNx{978-0-19-169367-0}
\showISSN{1095-9203}
\showDOI{%
\url{http://dx.doi.org/10.1093/acprof:oso/9780198567523.003.0011}}


\bibitem{kassambaraRstatixPipeFriendlyFramework2019}
{Alboukadel Kassambara}. 2019.
\newblock Rstatix: {{Pipe}}-{{Friendly Framework}} for {{Basic Statistical
  Tests}}.
\newblock   (dec 2019).
\newblock


\bibitem{khumsriPrevalenceFacebookAddiction2015}
{Jiraporn Khumsri}, {Rungmanee Yingyeun}, {null {Mereerat Manwong}}, {Nitt
  Hanprathet}, {and} {Muthita Phanasathit}. 2015.
\newblock \showarticletitle{Prevalence of {{Facebook Addiction}} and {{Related
  Factors Among Thai High School Students}}}.
\newblock {\em Journal of the Medical Association of Thailand = Chotmaihet
  Thangphaet\/}  {98 Suppl 3} (apr 2015), S51--60.
\newblock
\showISSN{0125-2208}


\bibitem{kimTechnologySupportedBehavior2017}
{Jaejeung Kim}, {Chiwoo Cho}, {and} {Uichin Lee}. 2017.
\newblock \showarticletitle{Technology {{Supported Behavior Restriction}} for
  {{Mitigating Self}}-{{Interruptions}} in {{Multi}}-Device {{Environments}}}.
\newblock {\em Proceedings of the ACM on Interactive, Mobile, Wearable and
  Ubiquitous Technologies\/} {1}, 3 (sep 2017), 1--21.
\newblock
\showISSN{24749567}
\showDOI{%
\url{http://dx.doi.org/10.1145/3130932}}


\bibitem{kocFacebookAddictionTurkish2013}
{Mustafa Koc} {and} {Seval Gulyagci}. 2013.
\newblock \showarticletitle{Facebook Addiction among {{Turkish}} College
  Students: The Role of Psychological Health, Demographic, and Usage
  Characteristics}.
\newblock {\em Cyberpsychology, Behavior and Social Networking\/} {16}, 4 (apr
  2013), 279--284.
\newblock
\showISSN{2152-2723}
\showDOI{%
\url{http://dx.doi.org/10.1089/cyber.2012.0249}}


\bibitem{kotabeIntegratingComponentsSelfControl2015}
{Hiroki~P. Kotabe} {and} {Wilhelm Hofmann}. 2015.
\newblock \showarticletitle{On {{Integrating}} the {{Components}} of
  {{Self}}-{{Control}}}.
\newblock {\em Perspectives on Psychological Science\/} {10}, 5 (2015),
  618--638.
\newblock
\showISSN{17456924}
\showDOI{%
\url{http://dx.doi.org/10.1177/1745691615593382}}


\bibitem{kovacsConservationProcrastinationProductivity2019}
{Geza Kovacs}, {Drew~Mylander Gregory}, {Zilin Ma}, {Zhengxuan Wu}, {Golrokh
  Emami}, {Jacob Ray}, {and} {Michael~S. Bernstein}. 2019.
\newblock \showarticletitle{Conservation of {{Procrastination}}: {{Do
  Productivity Interventions Save Time Or Just Redistribute It}}?}. In {\em
  Proceedings of the 2019 {{CHI Conference}} on {{Human Factors}} in
  {{Computing Systems}}} {\em ({{CHI}} '19)}. {ACM}, {New York, NY, USA},
  330:1--330:12.
\newblock
\showISBNx{978-1-4503-5970-2}
\showDOI{%
\url{http://dx.doi.org/10.1145/3290605.3300560}}


\bibitem{Kovacs2018}
{Geza Kovacs}, {Zhengxuan Wu}, {and} {Michael~S. Bernstein}. 2018.
\newblock \showarticletitle{Rotating Online Behavior Change Interventions
  Increases Effectiveness But Also Increases Attrition}.
\newblock {\em Proceedings of the {ACM} on Human-Computer Interaction\/} {2},
  {CSCW} (Nov. 2018), 1--25.
\newblock
\showDOI{%
\url{http://dx.doi.org/10.1145/3274364}}


\bibitem{krasnovaEnvyFacebookHidden2013}
{Hanna Krasnova}, {Helena Wenninger}, {Thomas Widjaja}, {and} {Peter Buxmann}.
  2013.
\newblock \showarticletitle{Envy on {{Facebook}}: {{A Hidden Threat}} to
  {{Users}}' {{Life Satisfaction}}?}. In {\em Wirtschaftsinformatik}.
\newblock
\showDOI{%
\url{http://dx.doi.org/10.7892/boris.47080}}


\bibitem{labragueFacebookUseAdolescents2014}
{Leodoro~J. Labrague}. 2014.
\newblock \showarticletitle{Facebook Use and Adolescents' Emotional States of
  Depression, Anxiety, and Stress}.
\newblock {\em Health Science Journal\/} {8}, 1 (2014), 80--89.
\newblock
\showISSN{1108-7366}


\bibitem{avtechlabsAutoLogout2019}
{AV~Tech Labs}. 2019.
\newblock Auto {{Logout}}.
\newblock   (jul 2019).
\newblock


\bibitem{lascauWhyAreCrossDevice2019}
{Laura Lascau}, {Priscilla N~Y Wong}, {Duncan~P Brumby}, {and} {Anna~L Cox}.
  2019.
\newblock Why {{Are Cross}}-{{Device Interactions Important When It Comes To
  Digital Wellbeing}}?  (2019).
\newblock


\bibitem{Lee2011Mining}
{Min~Kyung Lee}, {Sara Kiesler}, {and} {Jodi Forlizzi}. 2011.
\newblock \showarticletitle{Mining Behavioral Economics to Design Persuasive
  Technology for Healthy Choices}. In {\em Proceedings of the SIGCHI Conference
  on Human Factors in Computing Systems} {\em (CHI '11)}. ACM, New York, NY,
  USA, 325--334.
\newblock
\showISBNx{978-1-4503-0228-9}
\showDOI{%
\url{http://dx.doi.org/10.1145/1978942.1978989}}


\bibitem{lyngsTellMeWhat2018}
{Ulrik Lyngs}, {Reuben Binns}, {Max {Van Kleek}}, {and} {Nigel Shadbolt}. 2018.
\newblock \showarticletitle{{"}{{So}}, {{Tell Me What Users Want}}, {{What They
  Really}}, {{Really Want}}!{"}}. In {\em Extended {{Abstracts}} of the 2018
  {{CHI Conference}} on {{Human Factors}} in {{Computing Systems}}} {\em ({{CHI
  EA}} '18)}. {ACM}, {New York, NY, USA}, alt04:1----alt04:10.
\newblock
\showISBNx{978-1-4503-5621-3}
\showDOI{%
\url{http://dx.doi.org/10.1145/3170427.3188397}}


\bibitem{lyngsSelfControlCyberspaceApplying2019}
{Ulrik Lyngs}, {Kai Lukoff}, {Petr Slovak}, {Reuben Binns}, {Adam Slack},
  {Michael Inzlicht}, {Max {Van Kleek}}, {and} {Nigel Shadbolt}. 2019.
\newblock \showarticletitle{Self-{{Control}} in {{Cyberspace}} : {{Applying
  Dual Systems Theory}} to a {{Review}} of {{Digital Self}}-{{Control Tools}}}.
  In {\em {{CHIConference}} on {{Human Factors}} in {{Computing Systems
  Proceedings}} ({{CHI}} 2019)}. {ACM}, New York, NY, USA.
\newblock
\showDOI{%
\url{http://dx.doi.org/10.1145/3290605.3300361}}


\bibitem{marinoAssociationsProblematicFacebook2018}
{Claudia Marino}, {Gianluca Gini}, {Alessio Vieno}, {and} {Marcantonio~M.
  Spada}. 2018a.
\newblock \showarticletitle{The Associations between Problematic {{Facebook}}
  Use, Psychological Distress and Well-Being among Adolescents and Young
  Adults: {{A}} Systematic Review and Meta-Analysis}.
\newblock {\em Journal of Affective Disorders\/} {226}, September 2017 (2018),
  274--281.
\newblock
\showISSN{15732517}
\showDOI{%
\url{http://dx.doi.org/10.1016/j.jad.2017.10.007}}


\bibitem{marinoComprehensiveMetaanalysisProblematic2018}
{Claudia Marino}, {Gianluca Gini}, {Alessio Vieno}, {and} {Marcantonio~M
  Spada}. 2018b.
\newblock \showarticletitle{A Comprehensive Meta-Analysis on {{Problematic
  Facebook Use}}}.
\newblock {\em Computers in Human Behavior\/}  {83} (2018), 262--277.
\newblock
\showISSN{07475632}
\showDOI{%
\url{http://dx.doi.org/10.1016/j.chb.2018.02.009}}


\bibitem{markEffectsIndividualDifferences2018}
{Gloria Mark}, {Mary Czerwinski}, {and} {Shamsi~T Iqbal}. 2018.
\newblock \showarticletitle{Effects of {{Individual Differences}} in {{Blocking
  Workplace Distractions}}}. In {\em Proceedings of the 2018 {{CHI Conference}}
  on {{Human Factors}} in {{Computing Systems}}} {\em ({{CHI}} '18)}. {ACM},
  {New York, NY, USA}, 92:1----92:12.
\newblock
\showISBNx{978-1-4503-5620-6}
\showDOI{%
\url{http://dx.doi.org/10.1145/3173574.3173666}}


\bibitem{meierFacebocrastinationPredictorsUsing2016}
{Adrian Meier}, {Leonard Reinecke}, {and} {Christine~E. Meltzer}. 2016.
\newblock \showarticletitle{Facebocrastination? {{Predictors}} of {{Using
  Facebook}} for {{Procrastination}} and {{Its Effects}} on {{Students
  Well}}-Being}.
\newblock {\em Comput. Hum. Behav.\/} {64}, C (nov 2016), 65--76.
\newblock
\showISSN{0747-5632}
\showDOI{%
\url{http://dx.doi.org/10.1016/j.chb.2016.06.011}}


\bibitem{miriEmotionRegulation2018}
{Pardis Miri}, {Andero Uusberg}, {Heather Culbertson}, {Robert Flory}, {Helen
  Uusberg}, {James~J. Gross}, {Keith Marzullo}, {and} {Katherine Isbister}.
  2018.
\newblock \showarticletitle{Emotion Regulation in the Wild: Introducing WEHAB
  System Architecture}. In {\em Extended Abstracts of the 2018 CHI Conference
  on Human Factors in Computing Systems} {\em (CHI EA '18)}. ACM, New York, NY,
  USA, Article LBW021, 6 pages.
\newblock
\showISBNx{978-1-4503-5621-3}
\showDOI{%
\url{http://dx.doi.org/10.1145/3170427.3188495}}


\bibitem{mosqueraEconomicEffectsFacebook2019}
{Roberto Mosquera}, {Mofioluwasademi~(Moffii) Odunowo}, {Trent McNamara},
  {Xiongfei Guo}, {and} {Ragan Petrie}. 2019.
\newblock {\em The {{Economic Effects}} of {{Facebook}}}.
\newblock {{SSRN Scholarly Paper}} ID 3312462. {Social Science Research
  Network}, Rochester, NY.
\newblock


\bibitem{nicholsGoodSubjectEffectInvestigating2008}
{Austin~Lee Nichols} {and} {Jon~K. Maner}. 2008.
\newblock \showarticletitle{The {{Good}}-{{Subject Effect}}: {{Investigating
  Participant Demand Characteristics}}}.
\newblock {\em The Journal of General Psychology\/} {135}, 2 (apr 2008),
  151--166.
\newblock
\showISSN{0022-1309}
\showDOI{%
\url{http://dx.doi.org/10.3200/GENP.135.2.151-166}}


\bibitem{orbenScreensTeensPsychological2019}
{Amy Orben} {and} {Andrew~K. Przybylski}. 2019.
\newblock \showarticletitle{Screens, {{Teens}}, and {{Psychological
  Well}}-{{Being}}: {{Evidence From Three Time}}-{{Use}}-{{Diary Studies}}}.
\newblock {\em Psychological Science\/} {30}, 5 (may 2019), 682--696.
\newblock
\showISSN{1467-9280}
\showDOI{%
\url{http://dx.doi.org/10.1177/0956797619830329}}


\bibitem{oroszFourFacetsFacebook2016}
{G{\a'a}bor Orosz}, {Istv{\a'a}n T{\a'o}th-Kir{\a'a}ly}, {and} {Be{\a'a}ta B{\H
  o}the}. 2016.
\newblock \showarticletitle{Four Facets of {{Facebook}} Intensity \textemdash{}
  {{The}} Development of the {{Multidimensional Facebook Intensity Scale}}}.
\newblock {\em Personality and Individual Differences\/}  {100} (oct 2016),
  95--104.
\newblock
\showISSN{0191-8869}
\showDOI{%
\url{http://dx.doi.org/10.1016/j.paid.2015.11.038}}


\bibitem{PerezVallejos2017}
{Elvira Perez~Vallejos}, {Ansgar Koene}, {Virginia Portillo}, {Liz Dowthwaite},
  {and} {Monica Cano}. 2017.
\newblock \showarticletitle{Young People's Policy Recommendations on Algorithm
  Fairness}. In {\em Proceedings of the 2017 ACM on Web Science Conference}
  {\em (WebSci '17)}. ACM, New York, NY, USA, 247--251.
\newblock
\showISBNx{978-1-4503-4896-6}
\showDOI{%
\url{http://dx.doi.org/10.1145/3091478.3091512}}


\bibitem{phuFacebookUseIts2019}
{Becky Phu} {and} {Alan~J. Gow}. 2019.
\newblock \showarticletitle{Facebook Use and Its Association with Subjective
  Happiness and Loneliness}.
\newblock {\em Computers in Human Behavior\/}  {92} (mar 2019), 151--159.
\newblock
\showDOI{%
\url{http://dx.doi.org/10.1016/j.chb.2018.11.020}}


\bibitem{pinderDigitalBehaviourChange2018}
{Charlie Pinder}, {Jo Vermeulen}, {Benjamin~R. Cowan}, {and} {Russell Beale}.
  2018.
\newblock \showarticletitle{Digital {{Behaviour Change Interventions}} to
  {{Break}} and {{Form Habits}}}.
\newblock {\em ACM Transactions on Computer-Human Interaction\/} {25}, 3
  (2018), 1--66.
\newblock
\showISSN{10730516}
\showDOI{%
\url{http://dx.doi.org/10.1145/3196830}}


\bibitem{pollerWelcomeROSEResearch2017}
{Andreas Poller}. 2017.
\newblock Welcome to {{ROSE}} - {{The Research Tool}} for {{Online Social
  Environments}}.
\newblock   (2017).
\newblock


\bibitem{pollerInvestigatingOSNUsers2014}
{Andreas Poller}, {Petra Ilyes}, {Andreas Kramm}, {and} {Laura Kocksch}. 2014.
\newblock \showarticletitle{Investigating {{OSN Users}}' {{Privacy Strategies}}
  with {{In}}-Situ {{Observation}}}. In {\em Proceedings of the {{Companion
  Publication}} of the 17th {{ACM Conference}} on {{Computer Supported
  Cooperative Work}} \& {{Social Computing}}} {\em ({{CSCW Companion}} '14)}.
  {ACM}, {New York, NY, USA}, 217--220.
\newblock
\showISBNx{978-1-4503-2541-7}
\showDOI{%
\url{http://dx.doi.org/10.1145/2556420.2556508}}


\bibitem{prettymind.coTimewasteTimer2018}
{{Prettymind.co}}. 2018.
\newblock {\em Timewaste {{Timer}}}.
\newblock


\bibitem{przybylskiMotivationalEmotionalBehavioral2013}
{Andrew~K. Przybylski}, {Kou Murayama}, {Cody~R. Dehaan}, {and} {Valerie
  Gladwell}. 2013.
\newblock \showarticletitle{Motivational, Emotional, and Behavioral Correlates
  of Fear of Missing Out}.
\newblock {\em Computers in Human Behavior\/} {29}, 4 (2013), 1841--1848.
\newblock
\showISSN{07475632}
\showDOI{%
\url{http://dx.doi.org/10.1016/j.chb.2013.02.014}}


\bibitem{redelmeierPatientsMemoriesPainful1996}
{Donald~A. Redelmeier} {and} {Daniel Kahneman}. 1996.
\newblock \showarticletitle{Patients' Memories of Painful Medical Treatments:
  {{Real}}-Time and Retrospective Evaluations of Two Minimally Invasive
  Procedures}.
\newblock {\em Pain\/} {66}, 1 (1996), 3--8.
\newblock
\showISSN{03043959}
\showDOI{%
\url{http://dx.doi.org/10.1016/0304-3959(96)02994-6}}


\bibitem{reineckeSlackingWindingExperience2016}
{Leonard Reinecke} {and} {Wilhelm Hofmann}. 2016.
\newblock \showarticletitle{Slacking off or {{Winding}} down? {{An Experience
  Sampling Study}} on the {{Drivers}} and {{Consequences}} of {{Media Use}} for
  {{Recovery}} versus {{Procrastination}}}.
\newblock {\em Human Communication Research\/} {42}, 3 (jul 2016), 441--461.
\newblock
\showISSN{0360-3989}
\showDOI{%
\url{http://dx.doi.org/10.1111/hcre.12082}}


\bibitem{robinsMeasuringGlobalSelfesteem2001}
{Richard~W. Robins}, {Holly~M. Hendin}, {and} {Kali~H. Trzesniewski}. 2001.
\newblock \showarticletitle{Measuring Global Self-Esteem: {{Construct}}
  Validation of a Single-Item Measure and the {{Rosenberg Self}}-{{Esteem
  Scale}}}.
\newblock {\em Personality and Social Psychology Bulletin\/} {27}, 2 (2001),
  151--161.
\newblock
\showISSN{1552-7433(Electronic),0146-1672(Print)}
\showDOI{%
\url{http://dx.doi.org/10.1177/0146167201272002}}


\bibitem{rosenFacebookTextingMade2013}
{Larry~D. Rosen}, {L. {Mark Carrier}}, {and} {Nancy~A. Cheever}. 2013.
\newblock \showarticletitle{Facebook and Texting Made Me Do It:
  {{Media}}-Induced Task-Switching While Studying}.
\newblock {\em Computers in Human Behavior\/} {29}, 3 (2013), 948--958.
\newblock
\showISSN{07475632}
\showDOI{%
\url{http://dx.doi.org/10.1016/j.chb.2012.12.001}}


\bibitem{rouisImpactFacebookUsage2011}
{Sana Rouis}, {Moez Limayem}, {and} {Esmail Salehi-Sangari}. 2011.
\newblock \showarticletitle{Impact of {{Facebook Usage}} on {{Students}}'
  {{Academic Achievement}}: {{Role}} of Self-Regulation and Trust}.
\newblock {\em Electronic Journal of Research in Educational Psychology\/} 25
  (2011), 34.
\newblock


\bibitem{ryanUsesAbusesFacebook2014}
{Tracii Ryan}, {Andrea Chester}, {John Reece}, {and} {Sophia Xenos}. 2014.
\newblock \showarticletitle{The Uses and Abuses of {{Facebook}}: {{A}} Review
  of {{Facebook}} Addiction}.
\newblock {\em Journal of Behavioural Addictions\/} {3}, 3 (2014), 133--148.
\newblock
\showISSN{2062-5871}
\showDOI{%
\url{http://dx.doi.org/10.1556/JBA.3.2014.016}}


\bibitem{scharkowAccuracySelfReportedInternet2016}
{Michael Scharkow}. 2016.
\newblock \showarticletitle{The {{Accuracy}} of {{Self}}-{{Reported Internet
  Use}} - {{A Validation Study Using Client Log Data}}}.
\newblock {\em Communication Methods and Measures\/} {10}, 1 (jan 2016),
  13--27.
\newblock
\showISSN{1931-2458}
\showDOI{%
\url{http://dx.doi.org/10.1080/19312458.2015.1118446}}


\bibitem{Swim2013}
{Janet~K. Swim} {and} {Brittany Bloodhart}. 2013.
\newblock \showarticletitle{Admonishment and Praise: Interpersonal Mechanisms
  for Promoting Proenvironmental Behavior}.
\newblock {\em Ecopsychology\/} {5}, 1 (March 2013), 24--35.
\newblock
\showDOI{%
\url{http://dx.doi.org/10.1089/eco.2012.0065}}


\bibitem{ticeEmotionalDistressRegulation2001}
{D~M Tice}, {E Bratslavsky}, {and} {R~F Baumeister}. 2001.
\newblock \showarticletitle{Emotional Distress Regulation Takes Precedence over
  Impulse Control: If You Feel Bad, Do It!}
\newblock {\em Journal of personality and social psychology\/} {80}, 1 (2001),
  53--67.
\newblock
\showISSN{0022-3514}
\showDOI{%
\url{http://dx.doi.org/10.1037/0022-3514.80.1.53}}


\bibitem{tromholtFacebookExperimentQuitting2016}
{Morten Tromholt}. 2016.
\newblock \showarticletitle{The {{Facebook Experiment}}: {{Quitting Facebook
  Leads}} to {{Higher Levels}} of {{Well}}-{{Being}}}.
\newblock {\em Cyberpsychology, Behavior and Social Networking\/} {19}, 11 (nov
  2016), 661--666.
\newblock
\showISSN{2152-2723}
\showDOI{%
\url{http://dx.doi.org/10.1089/cyber.2016.0259}}


\bibitem{OptimizingEngagementUnderstanding2019}
{{U.S. Senate Committee On Commerce, Science, \& Transportation}}. 2019.
\newblock Optimizing for {{Engagement}}: {{Understanding}} the {{Use}} of
  {{Persuasive Technology}} on {{Internet Platforms}}.
\newblock   (jun 2019).
\newblock
\showURL{%
\url{https://www.commerce.senate.gov/public/index.cfm/2019/6/optimizing-for-engagement-understanding-the-use-of-persuasive-technology-on-internet-platforms}}


\bibitem{vanmanBurdenOnlineFriends2018}
{Eric~J. Vanman}, {Rosemary Baker}, {and} {Stephanie~J. Tobin}. 2018.
\newblock \showarticletitle{The Burden of Online Friends: {{The}} Effects of
  Giving up {{Facebook}} on Stress and Well-Being}.
\newblock {\em The Journal of Social Psychology\/} {158}, 4 (jul 2018),
  496--508.
\newblock
\showISSN{0022-4545}
\showDOI{%
\url{http://dx.doi.org/10.1080/00224545.2018.1453467}}


\bibitem{verduynPassiveFacebookUsage2015}
{Philippe Verduyn}, {David~Seungjae Lee}, {Jiyoung Park}, {Holly Shablack},
  {Ariana Orvell}, {Joseph Bayer}, {Oscar Ybarra}, {John Jonides}, {and} {Ethan
  Kross}. 2015.
\newblock \showarticletitle{Passive {{Facebook}} Usage Undermines Affective
  Well-Being: {{Experimental}} and Longitudinal Evidence}.
\newblock {\em Journal of Experimental Psychology. General\/} {144}, 2 (apr
  2015), 480--488.
\newblock
\showISSN{1939-2222}
\showDOI{%
\url{http://dx.doi.org/10.1037/xge0000057}}


\bibitem{wangContextCollegeStudents2018}
{Yiran Wang} {and} {Gloria Mark}. 2018.
\newblock \showarticletitle{The {{Context}} of {{College Students}}' {{Facebook
  Use}} and {{Academic Performance}}: {{An Empirical Study}}}. In {\em
  Proceedings of the 2018 {{CHI Conference}} on {{Human Factors}} in
  {{Computing Systems}}} {\em ({{CHI}} '18)}. {ACM}, {New York, NY, USA},
  418:1--418:11.
\newblock
\showISBNx{978-1-4503-5620-6}
\showDOI{%
\url{http://dx.doi.org/10.1145/3173574.3173992}}


\bibitem{wolniczakAssociationFacebookDependence2013}
{Isabella Wolniczak}, {Jos{\a'e}~Alonso C{\a'a}ceres-DelAguila}, {Gabriela
  Palma-Ardiles}, {Karen~J. Arroyo}, {Rodrigo Sol\'is-Visscher}, {Stephania
  Paredes-Yauri}, {Karina Mego-Aquije}, {and} {Antonio Bernabe-Ortiz}. 2013.
\newblock \showarticletitle{Association between {{Facebook}} Dependence and
  Poor Sleep Quality: A Study in a Sample of Undergraduate Students in
  {{Peru}}}.
\newblock {\em PloS One\/} {8}, 3 (2013), e59087.
\newblock
\showISSN{1932-6203}
\showDOI{%
\url{http://dx.doi.org/10.1371/journal.pone.0059087}}


\bibitem{xuMediaMultitaskingWellbeing2016}
{Shan Xu}, {Zheng~(Joyce) Wang}, {and} {Prabu David}. 2016.
\newblock \showarticletitle{Media {{Multitasking}} and {{Well}}-Being of
  {{University Students}}}.
\newblock {\em Comput. Hum. Behav.\/} {55}, PA (feb 2016), 242--250.
\newblock
\showISSN{0747-5632}
\showDOI{%
\url{http://dx.doi.org/10.1016/j.chb.2015.08.040}}


\end{thebibliography}

\end{document}